\documentclass[twocolumn]{aastex631}

\usepackage{graphicx}	
\usepackage{amsmath}	
\usepackage{color}

\def\hmpc{h^{-1}{\rm Mpc}}

\def\hkpc{h^{-1}\, {\rm kpc}}

\def\hmsun{{h^{-1} M_{\odot}}}
\def\msun{\, M_{\odot}}
\def\mbh{\, M_{\rm BH}}

\def\astrid{\texttt{ASTRID} }
\def\astridN{\texttt{ASTRID}}

\begin{document}

\title[ASTRID simulation]{The Astrid Simulation: Evolution of black holes and galaxies to $z=0.5$ and different evolution pathways for galaxy quenching}

%

\author{Yueying Ni}
\affiliation{Center for Astrophysics $\vert$ Harvard \& Smithsonian, 60 Garden St, Cambridge, MA 02138, USA}

\author{Nianyi Chen}
\affiliation{McWilliams Center for Cosmology, Department of Physics, Carnegie Mellon University, Pittsburgh, PA 15213, USA}
\affiliation{School of Natural Sciences, Institute for Advanced Study, Princeton, NJ 08540, USA}

\author{Yihao Zhou}
\affiliation{McWilliams Center for Cosmology, Department of Physics, Carnegie Mellon University, Pittsburgh, PA 15213, USA}

\author{Minjung Park}
\affiliation{Center for Astrophysics $\vert$ Harvard \& Smithsonian, 60 Garden St, Cambridge, MA 02138, USA}

\author{Yanhui Yang}
\affiliation{Department of Physics \& Astronomy, University of California, Riverside, 900 University Ave., Riverside, CA 92521, USA}

\author{Tiziana DiMatteo}
\affiliation{McWilliams Center for Cosmology, Department of Physics, Carnegie Mellon University, Pittsburgh, PA 15213, USA}

\author{Simeon Bird}
\affiliation{Department of Physics \& Astronomy, University of California, Riverside, 900 University Ave., Riverside, CA 92521, USA}

\author{Rupert Croft}
\affiliation{McWilliams Center for Cosmology, Department of Physics, Carnegie Mellon University, Pittsburgh, PA 15213, USA}


\begin{abstract}

We present new results from the \astrid simulation from $z=3$ to $z=0.5$, covering the epoch of cosmic noon.
The galaxy stellar mass function, as well as the black hole mass and luminosity functions in \astridN, exhibit good agreement with recent observational constraints. 
We study the $\mbh$-$M_*$ scaling relation and its connections to AGN luminosity, galaxy color, and star formation rate, demonstrating that AGN feedback plays a crucial role in the quenching of massive galaxies ($M_*>10^{10.5} \msun$). 
Although AGN feedback suppresses star formation through quenching, AGN-host galaxies still exhibit statistically higher levels of star formation compared to inactive ones, due to the positive correlation between AGN activity and star formation, both fueled by a shared gas reservoir. 
The fraction of quiescent galaxies in \astrid increases with both galaxy mass and redshift evolution, aligning well with observational trends. 
We find that different quenching mechanisms can leave distinct morphological imprints on quenched galaxies. 
Massive, compact quiescent galaxies typically experience shorter quenching timescales, have younger central regions, and host overmassive black holes. 
This is usually due to a compaction-like quenching mechanism that funnels gas into the galaxy center, leading to starbursts and triggering AGN kinetic feedback.
In contrast, quiescent galaxies with more diffuse morphologies generally experience `inside-out' quenching, which is characterized by older central regions compared to the outskirts. 
These galaxies typically experience longer quenching timescales due to quenching processes operating on a larger halo scale, which gradually deplete the galactic star-forming gas.
Data of the \astrid simulation down to $z=0.5$ is available at \url{https://astrid.psc.edu}.

\end{abstract}




\section{Introduction}
\label{sec:intro}

Modern large-scale cosmological hydrodynamical simulations have been highly effective in creating realistic galaxy and massive black hole populations across cosmic history.
Over the last decade, cosmological-scale hydrodynamic simulations, such as Magneticum~\citep{Hirschmann2014}, Illustris~\citep{Vogelsberger:2014}, Eagle~\citep{Schaye2015}, Horizon-AGN~\citep{Dubois2015,Volonteri2016Horizon-AGN}, MassiveBlack~\citep{Khandai2015}, BlueTides~\citep{Feng2016}, Romulus~\citep{Tremmel2017}, Illustris-TNG~\citep{Springel2018}, Simba~\citep{Dave2019-simba}, THESAN \citep{Kannan2022MNRAS.511.4005K}, FLAMINGO \citep{Schaye2023MNRAS.526.4978S} have implemented ever improving sub-grid models, and have made significant progress in modelling the galaxy and massive black hole evolution over cosmic time. 

This study follows on from the high-z evolution \citep{Ni-Asterix, Bird-Asterix} and introduces the low-$z$ \astrid results, in which we seek to make progress on pushing hydrodynamical simulations of galaxy formation to much larger volume at high resolution to study the evolution of galaxy and black hole populations. 

In particular, in this work, we present new results from the \astrid simulation from $z=3$ to $z=0.5$, giving an overview of galaxy and black hole evolution over the epoch of cosmic noon. In particular, we focus on the rise of the quiescent galaxy population. 
\astrid is a large volume, high resolution, cosmological hydrodynamic simulation, first introduced in \citep{Bird-Asterix,Ni-Asterix}. 
With a box of $250 \hmpc$ containing $2\times5500^3$ particles, the large dynamic range of \astrid enables detailed studies of galaxy and massive black hole evolution at high resolution while allowing systematic studies of rare massive systems within a large cosmic volume. 
So far, \astrid has consumed 670M CPU hours to reach $z=0.5$, making it one of the most computationally ambitious cosmological galaxy formation simulations to date. 

High-$z$ ($z>2-3$) results from the \astrid simulation have been used for several studies, including the re-ionization and galaxy population \citep{Bird-Asterix}, evolution of super-massive black holes \citep{Ni-Asterix}, black hole coalescence and gravitational waves \citep{Chen2022MNRAS.514.2220C,degraf24}, populations of dual AGN and their host galaxies \citep{Chen2022arXiv220804970C,shen2023,dadiani23}, star-forming linear wakes in galaxy encounters \citep{flyby}, ultramassive black holes and quasar triplets \citep{Ni2022ApJ...940L..49N,Hoffman2023MNRAS.524.1987H}, intermediate mass and wandering black holes \citep{dimatteo2023,Weller23}, galaxy morphology and mock predictions for JWST \citep{LaChance2024arXiv240116608L}, the topology and statistics of the Epoch of Reionization \citep{Davies:2023}, line intensity mapping and Lyman-alpha forest\citep{2023MNRAS.524.1933Q}, seed black holes and their orbits \citep{magics1}.

Cosmic star formation and AGN activity peak at $z \sim 2$, an epoch often referred to as cosmic noon, and gradually decline thereafter \citep{Madau2014ARA&A..52..415M}.
It has been known for several decades that galaxies in the current universe are divided into two types: star-forming and quiescent, leading to the observed color bimodality of blue and red galaxies in large galaxy surveys \citep[e.g.,][]{Strateva2001AJ....122.1861S,Baldry2004ApJ...600..681B}. 
While most massive galaxies are star-forming at $z \sim 2$, a population of quiescent galaxies builds up over cosmic time and dominates the massive end of the galaxy stellar mass function at $z \sim 0$ \citep[][]{Ilbert2013A&A...556A..55I, Muzzin2013ApJ...777...18M, Davidzon2017A&A...605A..70D}. 
This implies that during and after the epoch of cosmic noon, a significant number of galaxies transition from star-forming to quiescent, a process often referred to as `quenching'.

Many different quenching mechanisms have been proposed \citep[e.g.,][]{Man2018NatAs...2..695M}, which can be roughly categorized by different spatial scales and time scales.

By spatial scales, quenching mechanisms can be roughly divided into two types: `internal' and `external' \citep[or  ``mass'' and ``environmental'' quenching, as proposed in ][]{Peng2010ApJ...721..193P}.
Internal or central quenching mechanisms arise from physical processes within the galaxy, such as ejective AGN feedback \citep{Silk1998A&A...331L...1S}, morphological quenching from the galaxy bulge/bar or disk instabilities \citep[e.g.,][]{Martig2009ApJ...707..250M}, etc. 
External or environmental quenching mechanisms operate on larger scales than the host halo, such as heating of gas within halos \citep[e.g.,][]{Croton2006MNRAS.365...11C}, ram pressure stripping \citep[e.g.,][]{Gunn1972ApJ...176....1G}, and gas starvation \citep[e.g.,][]{Feldmann2015MNRAS.446.1939F,Trussler2020MNRAS.491.5406T}. 
Environmental quenching accounts for the abundance of quiescent lower-mass systems in large-scale overdense regions and is the dominant quenching mechanism for low-mass satellite galaxies \citep[e.g.,][]{Jung2018ApJ...865..156J}.

By time scales, 
observations have identified a wide range of galaxy quenching timescales, typically categorized as “slow” and “rapid” quenching \citep[e.g.,][]{Schawinski2014MNRAS.440..889S, Yesuf2014ApJ...792...84Y, Barro2016ApJ, Wu2018ApJ...868...37W, Belli2019ApJ...874...17B, Carnall2019, Suess2021ApJ...915...87S,Tacchella2022ApJ,Park2023ApJ...953..119P}, indicating distinct evolutionary pathways of quenching.

Since different quenching mechanisms operate on varying spatial and time scales, they can also leave distinct imprints on galaxy morphological properties such as compactness and color gradient (i.e., the stellar age profile). 
Many observational studies have been conducted to unveil the spectroscopic and morphological details of the quiescent galaxy population at both low ($z \sim 0$) and high redshifts ($z>1$). 
Those studies aim to provide insights into galaxy quenching mechanisms by reconstructing star formation histories (SFH), constraining quenching epochs and timescales, and investigating the correlation between quenching and galaxy morphology.

For example, based on the local quiescent galaxy (QG) population, \cite{Schawinski2014MNRAS.440..889S} find that rapidly quenched galaxies are associated with early-type bulge spheroidal, while slowly quenched galaxies are linked with late-type disky morphologies. 
\cite{Belli2019ApJ...874...17B} investigated a population of $1.5<z<2.5$ galaxies and found that rapid quenching produces compact post-starburst systems, whereas slow quenching leads to galaxies of larger sizes.

Different quenching evolution pathways can also result in distinct galaxy age gradients. 
Observations of low-$z$ galaxies \citep{Woo2019MNRAS.487.1927W} and high-$z$ quiescent galaxies ($z>1$) \citep[e.g.,][]{Barro2016ApJ, Suess2021ApJ...915...87S} reveal that rapidly quenched galaxies undergo central compaction-like quenching, which involves a dissipative core-building event that can trigger a central starburst and fuel AGN, are more compact and exhibit positive color gradients (i.e., have bluer and younger stellar centers compared to the outskirts). 
In contrast, galaxies that undergo quenching unrelated to central processes tend to be larger and have older stellar centers, exhibiting negative color gradients.


In observations, high-redshift distant quiescent galaxies are much more compact than massive early-type galaxies in the present-day universe, making them more challenging to resolve.
However, with the advent of the James Webb Space Telescope (JWST) and its capability for spatially resolved deep spectroscopy, it will be possible to probe more detailed characteristics of the quiescent galaxy population around the epoch of cosmic noon ($1<z<3$) \citep[e.g.,][]{Slob2024arXiv240412432S, Park2024arXiv240417945P} and beyond, shedding light on the quenching mechanisms in the early universe.



In this work, we present the low-$z$ \astrid simulation results and inspect the emergence of the quiescent galaxy population through various pathways.
The paper is organized as follows. 
Section~\ref{Sec:method} provides a brief overview of the astrophysical models used in the \astrid simulation. 
In Section~\ref{Sec:galaxy-bh-population}, we show the results of some basic statistics of the galaxy and BH population in \astrid at $z=3$ to $z=0.5$. 
Section~\ref{sec:QG} focuses on the quiescent galaxy population in \astridN, exploring their detailed morphological properties and studying their correlations with quenching timescales. 
Finally, we summarize the paper in Section~\ref{Sec:summary}.

\section{ASTRID simulation}
\label{Sec:method}

The \astrid simulation is performed using a new version of the \texttt{MP-Gadget} simulation code \citep{MPGadget2018}, a massively scalable version of the cosmological structure formation code Gadget-3 \citep{Springel:2005}. 
The simulation contains $5500^3$ cold dark matter (DM) particles in a $250 \hmpc$ side box, and an initially equal number of SPH hydrodynamic mass elements. 
The cosmological parameters used are from \citep{Planck}, with $\Omega_0=0.3089$, $\Omega_\Lambda=0.6911$, $\Omega_{\rm b}=0.0486$, $h=0.6774$, $A_s = 2.142 \times 10^{-9}$ ($\sigma_8=0.816$), $n_s=0.9667$. 
The mass resolution of \astrid is $M_{\rm DM} = 6.74 \times 10^6 \hmsun$ and $M_{\rm gas} = 1.27 \times 10^6 \hmsun$ in the initial conditions. 
The gravitational softening length is $\epsilon_{\rm g} = 1.5 \hkpc$ for both DM and gas particles.
The initial conditions are set at $z=99$ and the current final redshift is $z=0.5$.

In this section, we briefly list some of the basic features of the astrophysical models in \astridN, and refer readers to earlier introductory papers for a more extensive description of the simulation code and implementations.

\subsection{Galaxy formation and reionization models}

In \astridN, gravitational dynamics are computed using the TreePM algorithm. 
For the hydrodynamics, we solve the Euler equations using the pressure-entropy formulation of smoothed particle hydrodynamics (pSPH)  \citep{Hopkins2013MNRAS.428.2840H}. 

Gas is allowed to cool radiatively through primordial gas following \cite{Katz1996ApJS..105...19K} and via metal line cooling. 
We approximate the metal cooling rate by scaling a solar metallicity template according to the metallicity of gas particles, following \cite{Vogelsberger:2014}. 

We model patchy reionization with a spatially varying ultra-violet background using a semi-analytic method based on hydrodynamic simulations performed with radiative transfer \citep[for more details see][]{Battaglia2013ApJ...776...81B}. 
Within ionized regions, we apply the ionizing ultra-violet background provided by \cite{FG2020}. 
Self-shielding is implemented following \cite{Rahmati:2013}, ensuring that gas is shielded from the ultra-violet background and thus is neutral at a density above $0.01$ atoms/cm$^{-3}$.
We also include models for (patchy) helium reionization and the gravitational effect of massive neutrinos.

Star formation is implemented based on the multi-phase star formation model in \cite{SH03} and accounts for the effects of molecular hydrogen using the $\rm H_2$ fraction calculated from the metallicity and local column density~\citep{Krumholz2011ApJ...729...36K}. 
Stars are formed with $1/4$ of the mass of a gas particle (i.e., $M_{*} \sim 3\times 10^5 \hmsun$ in most cases). 
Type II supernova wind feedback is included following \cite{Okamoto2010}, assuming wind speeds proportional to the local one dimensional dark matter velocity dispersion.
\astrid tracks metal enrichment from AGB stars, Type II SNe, and Type Ia SNe, following 9 individual elements (H, He, C, N, O, Ne, Mg, Si, Fe), following \cite{Vogelsberger2013}.

Galactic winds driven by stellar feedback are implemented kinetically via temporarily hydrodynamically decoupled particles.
The asymptotic mass loading factor of the SN wind scales with the wind speed by $\eta_w \propto v_w^{-2}$.
Winds are sourced by newly formed star particles, which randomly pick gas particles from within their SPH smoothing length to become wind particles. 
Once a particle is in the wind, it is hydrodynamically decoupled for the minimum of $60\,$Myr or $20\,\mathrm{kpc} / v_w$, and is recoupled once its density drops by a factor of $10$. 
Particles in the wind do not experience or produce pressure forces, but they do receive the mass return, cool, and contribute to density estimates. 


\subsection{BH models}

The supermassive black hole (SMBH) models in \astrid are described in \cite{Ni-Asterix}.
In this section, we briefly go over the modeling of BH seeding, accretion, and dynamics. We refer to \cite{Ni-Asterix} for a full description.
In particular, we note that \astrid has BH kinetic feedback enabled after $z=2.3$, as described in more detail in Section.~\ref{subsection:AGN-fdbk}. 

\subsubsection{BH seeding}

In \astridN, BHs are seeded within sufficiently massive halos. 
A Friends-of-Friends (FOF) group finder identifies eligible halos with a total mass exceeding $5 \times 10^9 \hmsun$ and stellar mass over $2 \times 10^6 \hmsun$. 
This ensures BHs form in halos with enough cold dense gas to form stars, and have collisionless star particles for dynamical friction.

Given the complexity of BH seed formation, \astrid uses a probabilistic approach for BH seed masses, drawing from a power-law distribution with $M_{\rm sd,min} = 3 \times 10^4 \hmsun$, $M_{\rm sd,max} = 3 \times 10^5 \hmsun$, and a power-law index $n = -1$.

For each halo that satisfies the seeding criteria but does not already contain at least one SMBH particle, we convert the densest gas particle into a BH particle. 
The initial BH seed mass, $M_{\rm BH}$, starts at $M_{\rm sd}$, with a separate mass variable tracking the parent particle's gas reservoir. 
Gas particles are accreted once $M_{\rm BH}$ grows beyond the initial parent particle mass.

\subsubsection{BH accretion}
\label{subsection:BH-accretion}

BHs in \astrid accrete mass from nearby gas particles. 
The accretion rate onto the BH, estimated using a Bondi-Hoyle-Lyttleton-like prescription \citep{DSH2005}, depends on the local properties of gas particles within the BH's vicinity:
\begin{equation}
\label{equation:Bondi}
    \dot{M}_{\rm B} = \frac{4 \pi \alpha G^2 M_{\rm BH}^2 \rho}{(c^2_s+v_{\rm rel}^2)^{3/2}}
\end{equation}
Here, $c_s$ and $\rho$ represent the local sound speed and density of gas, respectively, and $v_{\rm rel}$ is the relative velocity of the BH with respect to nearby gas. 
A dimensionless fudge parameter $\alpha = 100$ corrects for the underestimation of the accretion rate due to unresolved phases of the subgrid interstellar medium.

Super-Eddington accretion is allowed briefly in the simulation but is capped at twice the Eddington accretion rate:
\begin{equation}
\label{equation:Meddington}
    \dot{M}_{\rm Edd} = \frac{4 \pi G M_{\rm BH} m_\mathrm{p}}{\eta \sigma_\mathrm{T} c}\,.
\end{equation}
where $m_\mathrm{p}$ is the proton mass, $\sigma_\mathrm{T}$ is the Thompson cross section, $c$ is the speed of light, and $\eta = 0.1$ is the radiative efficiency. 

The total BH accretion rate is determined as::
\begin{equation}
    \dot{M}_{\rm BH} = {\rm Min} (\dot{M}_{\rm B}, 2\dot{M}_{\rm Edd})
\end{equation}

Numerically, the physical BH mass, $\mbh$, increases continuously over time, and we separately track the BH dynamic mass $M_{\rm dyn}$ by stochastic accretion of nearby gas particles with a probability that ensures the dynamic mass tracks the BH mass over time.

\subsubsection{BH dynamics and merger}

When an SMBH moves through a medium of smaller collisionless particles, it experiences a drag force called dynamical friction, caused by the gravitational wake from the perturbed particles \citep{Chandrasekhar1943}. 
This force dissipates the momentum of the SMBHs, causing their orbits to decay toward the gravitational potential minimum and stabilizing them at the galactic center.

We estimate dynamical friction on SMBHs using the implementation of \cite{Chen2021}:
\begin{equation}
    \label{eq:H14}
    \mathbf{F}_{\rm DF} = -4\pi \rho_{\rm sph} \left(\frac{GM_{\rm dyn}}{v_{\rm BH}}\right)^2  \;\text{log}(\Lambda) \mathcal{F}\left(\frac{v_{\rm BH}}{\sigma_v}\right) \frac{\bf{v}_{\rm BH}}{v_{\rm BH}}.
\end{equation}
where $M_{\rm dyn}$ is the BH dynamical mass,  $\textbf{v}_{\rm BH}$ is the BH velocity relative to its surrounding medium,  $\rho_{\rm sph}$ is the density of dark matter and star particles within the SPH kernel, and $\sigma_v$ is the velocity dispersion of the surrounding particles. 

The Coulomb logarithm $\text{log}(\Lambda)$ in Eq.~\ref{eq:H14} is
\begin{equation}
    \Lambda = \frac{b_{\rm max}}{(GM_{\rm dyn})/v_{\rm BH}^2}
\end{equation}
accounting for the effective range of the friction between the specified $b_{\rm min}$ and $b_{\rm max}$.
We choose $b_{\rm max} = 20\text{ kpc}$ as an ad hoc choice for the maximum physical range.

The function $\mathcal{F}$ in Eq.~\ref{eq:H14} is defined as
\begin{equation}
    \label{eq:fx}
    \mathcal{F}(x) =  \text{erf}(x)-\frac{2x}{\sqrt{\pi}} e^{-x^2}, \;
    x=\frac{v_{\rm BH}}{\sigma_v}
\end{equation}
This results from analytically integrating the surrounding particle velocity field $f(v_a)$, assuming a Maxwellian distribution \citep[as in, e.g.,][]{Binney2008}.

In \astridN, the BH seed mass is as low as $3 \times 10^4 \hmsun$, one order of magnitude smaller than the stellar particle mass. 
In this mass regime, dynamical friction is underestimated due to the limited mass resolution of the star particles, making the dynamics of the seed BHs unstable.
To stabilize BH motion during early post-seeding evolution, we use $M_{\rm dyn}$ instead of $M_{\rm BH}$ in Eq.~\ref{eq:H14}.

Dynamical friction modelling in \astrid results in well-defined BH orbits and velocities, allowing us to apply a more physical criterion for BH mergers based on the relative velocities and accelerations of merging SMBH pairs.
Following \cite{Bellovary2011} and \cite{Tremmel2017}, we determine if two BHs can be merged by evaluating their separation distance and determining whether they are gravitationally bound:
\begin{equation}
\label{equation:merger}
    \begin{cases}
    |\bf{\Delta r}| < 2 \epsilon_{\rm g} \\
    \frac{1}{2}|\bf{\Delta v}|^2 < \bf{\Delta a} \cdot \bf{\Delta r} \,.\\
   \end{cases}
\end{equation}
where $\bf{\Delta a}$, $\bf{\Delta v}$, and $\bf{\Delta r}$ are the relative acceleration, velocity, and position of the BH pair. 
$\epsilon_{\rm g} = 1.5 \hkpc$ is the gravitational softening length.
We set the merging distance based on $\epsilon_{\rm g}$ as BH dynamics below this distance are not spatially resolved.

This merging criterion improves over many cosmological simulations that use BH repositioning for BH dynamics, which can spuriously merge BHs with high relative velocities that are not gravitationally bound and should not merge yet (or may never merge). 
Detailed discussion of \astrid merger predictions can be found in \cite{Chen2021b}.

\begin{figure*}
\centering
\includegraphics[width=1.0\textwidth]{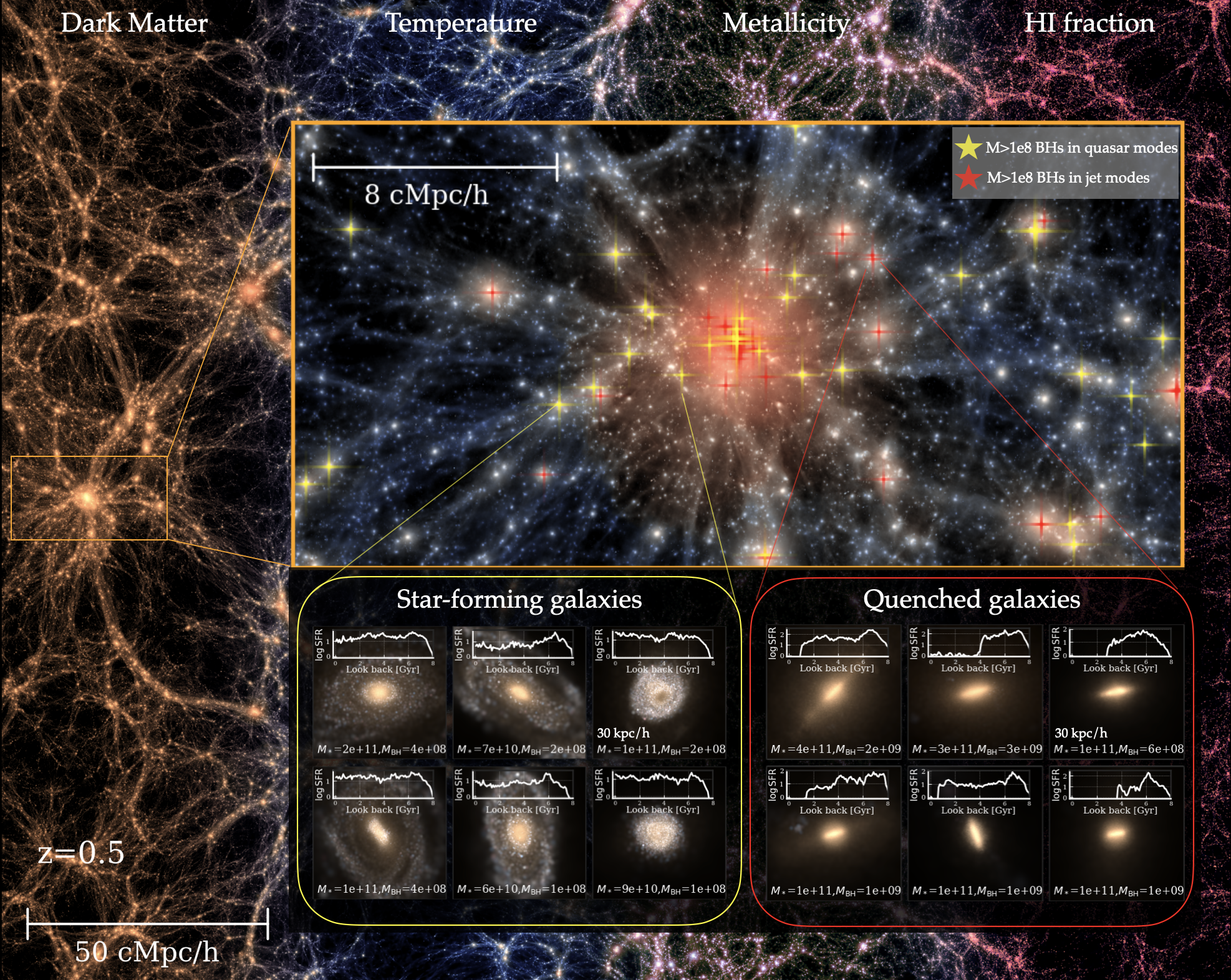}
\caption{Illustration of ASTRID at $z=0.5$. The background shows the cosmic web of size 250 $\hmpc$. The orange inset zooms into a massive cluster region, illustrating the gas density field colored by temperature. Red and yellow spikes mark all the massive black holes with $\mbh > 10^{8} \msun$; yellow indicates BHs in high-accretion quasar mode, typically residing in star-forming galaxies, while red indicates BHs in low-accretion jet mode, usually found in quiescent galaxies. The yellow and red insets show examples of the host galaxies within a region of 30 ckpc/h, with the RGB channels representing the flux in the rest-frame $grz$ color bands. The upper insets of each galaxy panel show the star formation history as a function of look-back time.
}
\label{fig:illustration}
\end{figure*}


\subsubsection{AGN feedback}
\label{subsection:AGN-fdbk}

AGN feedback plays a crucial role in regulating star formation and the growth of BHs. 
In most cosmological simulations, AGN feedback is the primary mechanism that quenches star formation in massive galaxies and explains the rise of the quiescent galaxy population after the epoch of cosmic noon.

AGN feedback can be broadly categorized into two regimes: radiatively efficient and radiatively inefficient. 
Radiatively efficient AGN dominate in the early universe when galaxies are gas-rich and actively forming stars. 
These AGN are often observed as quasars at high redshift, and are embedded in high-density environments with a high gas accretion rate, radiating a significant portion of the accretion energy as electromagnetic radiation. 
On the other side, radiatively inefficient AGN are more common in the low redshift universe.
In BHs with low Eddington accretion rates, the accretion flow is often advection-dominated, resulting in a lower radiative output compared to their radiatively efficient counterparts.
Those AGN primarily release energy through kinetic power carried by relativistic jets, which generate significant radio emission.
Thus, radiatively inefficient feedback is often referred to as jet or radio mode feedback \citep{Fabian2012ARA&A..50..455F}.

In \astridN, we implement radiatively efficient feedback (quasar mode) as thermal feedback, modeling the effect of radiative heating on the surrounding gas.
Radiatively inefficient feedback (jet mode) is implemented as kinetic feedback, effectively ejecting gas to quench star formation. 
The jet mode AGN feedback is activated at $z = 2.3$, corresponding to the time epoch when this feedback mechanism starts to play a significant role in galaxy evolution.

In \astridN, at $z>2.3$, AGN feedback operates solely in quasar mode (thermal feedback). 
After $z=2.3$, the AGN feedback follows a two-mode approach (1) thermal mode when $\lambda_{\rm Edd} > \chi_{\rm thr}$ and (2) kinetic mode when $\lambda_{\rm Edd} < \chi_{\rm thr}$;
The transition between thermal and kinetic feedback modes is determined by the Eddington ratio of the instantaneous BH accretion rate, 
with the Eddington threshold $\chi_{\rm thr}$ defined as:
\begin{equation}
    \chi_{\rm thr}=\min \left[\chi_{0}\left(\frac{M_{\mathrm{BH}}}{M_{\text{pivot}}}\right)^{\beta}, \chi_{\text{cap}}\right]
\end{equation}
where $\chi_0 = 0.002, \beta = 2, \chi_{\text{cap}} = 0.05, M_{\text{pivot}} = 5 \times 10^{8} \mathrm{M}_{\odot}$.
The Eddington threshold $\chi_{\rm thr}$ is capped at $\chi_{\text{thr,max}} = 0.05$ and depends on the BH mass, such that kinetic mode activates only for massive BHs with $\mbh \gtrsim 5 \times 10^{8} \hmsun$.
This AGN feedback model has been implemented and tested in \cite{Ni2023ApJ...959..136N}, demonstrating consistent galaxy properties compared to various observational constraints and other simulation results.

In the \textbf{high accretion mode}, the AGN feedback is deposited thermally:
\begin{equation}
\label{equation:thermalmode}
    \Delta \dot{E}_{\text {high }}= \epsilon_{\mathrm{f,th}} \epsilon_{\mathrm{r}} \dot{M}_{\mathrm{BH}} c^{2}   \,\,\,\,\, (\lambda_{\rm Edd} > \chi_{\rm thr}; \,\,\text{ thermal mode})
\end{equation}
Here we apply the mass-to-light conversion efficiency $\epsilon_{\mathrm r} = 0.1$ and $\epsilon_{\mathrm{f,th}} = 0.05$, assuming that 5\% of the radiation energy is thermally injected to the surrounding gas within a feedback sphere twice the SPH kernel radius. 
The AGN thermal feedback energy is isotropically imparted to the nearby gas particles, distributing the energy among them according to the SPH kernel weight.
Gas particles heated by AGN feedback dissipate energy based on their thermal properties: non-star-forming gas cools normally, while star-forming gas relaxes to the effective equation of state temperature on a cooling time scale.
We note that star-forming gas heated by other means cools to the effective equation of state on a longer relaxation timescale \citep{SH03}.

In the \textbf{low accretion mode}, the AGN feedback is deposited kinetically, following the prescription described in \citet{Weinberger2017} but with different parameter choices.
The AGN kinetic feedback energy is deposited as
\begin{equation}
   \Delta \dot{E}_{\text {low }} = \epsilon_{\mathrm{f, kin}} \dot{M}_{\mathrm{BH}} c^{2}  \,\,\,\,\, (\lambda_{\rm Edd} < \chi_{\rm thr};\,\,\text{jet mode})
\end{equation}
where the kinetic feedback efficiency $\epsilon_{\mathrm{f, kin}}$ is defined as:
\begin{equation}
    \epsilon_{\mathrm{f}, \text { kin }}=\min \left(\frac{\rho_{\text{BH}}}{f_{\text {thresh }} \rho_{\mathrm{SFthresh}}}, \epsilon_{\text{cap}}\right) 
\end{equation}
with $f_{\text {thresh }} = 0.05, \epsilon_{\text{cap}} = 0.05$.
Therefore, we have $\epsilon_{\mathrm{f, kin}}$ scaling with the BH local gas density and has a maximum value capped at $\epsilon_{\mathrm{f, kin,max}}=0.05$. 

In kinetic feedback mode, the feedback energy is accumulated over time and is released in a bursty way once the accumulated kinetic feedback energy exceeds the threshold $E_{\rm kin} > E_{\rm inj,min}$.
The minimum injected energy threshold $E_{\rm inj,min}$ is determined by
\begin{equation}
    E_{\mathrm{inj}, \mathrm{min}}=f_{\mathrm{re}} \frac{1}{2} \sigma_{\mathrm{DM}}^{2} m_{\mathrm{enc}}
\end{equation}
where $\sigma_{\mathrm{DM}}^{2}$ is the one-dimensional dark matter velocity dispersion around the BH, $m_{\mathrm{enc}}$ is the gas mass enclosed within the feedback kernel, and $f_{\mathrm{re}}=5$ controls the burstiness of the kinetic feedback. 
The released feedback energy $\Delta E_{\rm kin}$ kicks each gas particle within the feedback sphere in a random direction $\hat{n}$ with a prescribed momentum weighted by the SPH kernel:
\begin{equation}
    \Delta \boldsymbol{p}_{j}=m_{j} \sqrt{\frac{2 \Delta E_{\rm kin} w\left(\boldsymbol{r}_{j}\right)}{\rho}} \boldsymbol{\hat{n}}
\end{equation}
where $m_j$ and $w_j$ represents the mass and kernel weight of neighboring gas particles $j$ residing within the feedback sphere, and $\rho$ is the averaged gas density. 
Since the kick direction $\hat{n}$ is randomly chosen and changes each time the kinetic feedback energy is released, the kinetic mode feedback performs in an isotropic way when averaged over time.

\begin{figure*}
\centering
\includegraphics[width=1.0\textwidth]{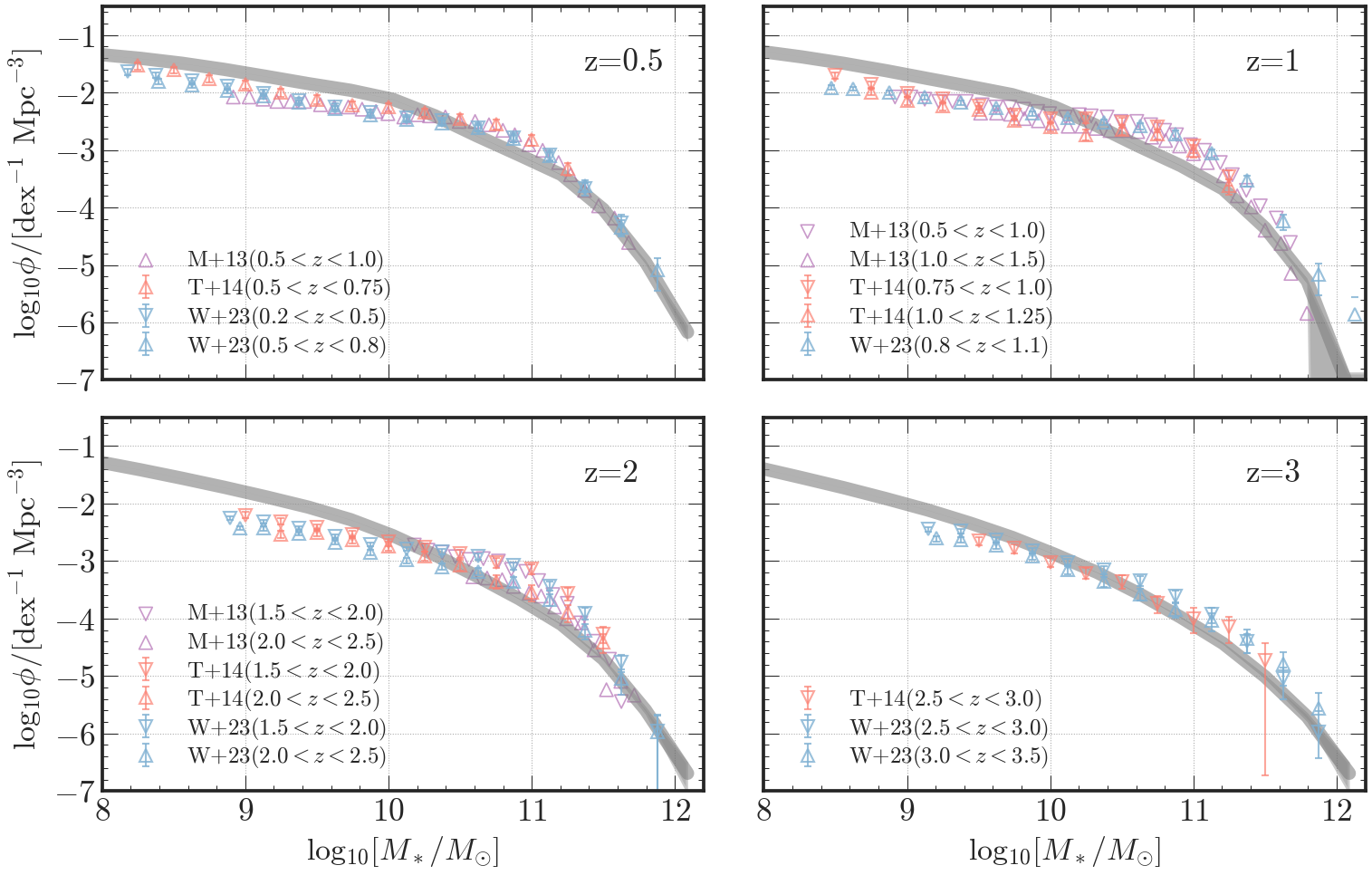}
\caption{The stellar mass function of \astrid from $z=3$ to $z=0.5$. For each galaxy, $M_{*}$ is calculated as the stellar mass enclosed within twice the half-mass radius of the associated subhalos, with this radius capped at 20kpc. Colored data points are from observational data from \citet{Muzzin2013ApJ...777...18M}(M+13), \citet{Tomczak2014ApJ...783...85T}(T+13) and \citet{Weaver2023A&A...677A.184W}(W+23). 
}
\label{fig:GSMF}
\end{figure*}

\section{Galaxy and black hole population}
\label{Sec:galaxy-bh-population}

Fig.~\ref{fig:illustration} provides an illustration of the \astrid simulation at $z=0.5$, visualizing a wide range of dynamical scales. The background shows a slice of the full simulation box, which is 250$\hmpc$ on each side (and a depth of 30$\hmpc$). 
We show a zoom into a cluster region of size $30 \times 15 \hmpc$ around one of the most massive halos with mass $M_h = 10^{15} \msun$, illustrating the gas density field colored by temperature. 
On larger scales, we can see that, consistent with observational findings \citep[e.g.,][]{Fabian2012ARA&A..50..455F}, at this low redshift regime ($z=0.5$), the most massive galaxies at the centers of clusters and groups generally do not host luminous AGN or quasars. Instead, these galaxies harbor the most massive SMBHs and are often active radio sources, operating in kinetic mode feedback involving jets acting on hot gas.
The lower panels illustrate various types of galaxy hosts for massive BHs. 
They show that most actively accreting BHs (in the quasar mode) are found in star-forming galaxies, while BHs in the jet mode are predominantly located in quiescent galaxies.
A detailed study of the quiescent galaxy population and their relation to BH activity is presented in Section~\ref{sec:QG}.

In this section, we start examining some overall statistics of the galaxy and black hole population of \astrid down to $z=0.5$ 
\citep[see also][for $z>3$ results]{Bird-Asterix, Ni-Asterix}.
In Section~\ref{sec3.1:gsmf} we focus on the galaxy stellar mass function, while in Section.~\ref{sec3.2:bhmf} we present some basic properties of the black hole population statistics. 
In Section.~\ref{sec3.3:scaling}, we show the scaling relation between galaxies and BHs, as well as the correlation between BH accretion and star-forming activities.

\subsection{Galaxy stellar mass function}
\label{sec3.1:gsmf}

We start with a comparison between the \astrid and observed galaxy stellar mass functions.
In \astridN, halos and subhalos are identified with the FOF and SUBFIND algorithm \citep{Springel2001MNRAS.328..726S}. 
To identify (sub)haloes, we first employ the standard FOF group finder with a linking length of 0.2 to identify FOF haloes. 
Subsequently, we apply the SUBFIND algorithm to these FOF groups to hierarchically identify substructures.
Throughout this work, we define galaxies with stellar mass threshold $M_{*} > 10^{8} \msun$, which corresponds to at least 300 star particles.

The galaxy stellar mass function (GSMF), defined as the number density of galaxies $\Phi(M,z)$ in bins of stellar mass $\Delta M$ at each redshift $z$, is a  fundamental cosmological observable to help us understand the formation and evolution of galaxies across cosmic history. 
Fig.~\ref{fig:GSMF} presents the results of the \astrid GSMF compared to a compendium of observational data \citep{Muzzin2013ApJ...777...18M, Tomczak2014ApJ...783...85T, Weaver2023A&A...677A.184W}.
In this analysis, $M_{*}$ is calculated as the stellar mass enclosed within 2 times the half mass radius of the associated subhalos for most galaxies.
However, in some massive halos ($M_h > 10^{14} \msun$), the central substructure often contains a diffuse stellar component that can sometime extend to hundreds of kpc.
This diffuse component results from the tidal stripping of stars from galaxies and the merging of smaller systems into the central cluster galaxy. 
To properly account for the mass of central massive galaxies in these clusters and align with observational data, we further apply a 3D spherical aperture with a radius of $R_p = 20$ ckpc/h. 
This approach excludes the stellar halo contribution, leading to the $M_{*}$ calculation defined as the stellar mass enclosed within $R_{\rm cut} = \min (R_{1/2,*}, R_p)$.
We note that this aperture cut only affects galaxies in the highest mass regime of $M_* \gtrsim 10^{12} \msun$, and changing the value of $R_p$ by a factor of 2 does not significantly change the overall statistics; most galactic radii are a few kpc. 

As shown in Fig.~\ref{fig:GSMF}, \astrid predicts the galaxy abundance in good agreement with observations over the entire mass range from $10^{8}$ to $10^{12} \msun$, including the redshift evolution.
At $z=2$ and $z=1$, the observational constraints exhibit a more pronounced `knee' in the mass range of $M_* \sim 10^{10.5-11.2} \msun$ compared to \astrid prediction, which shows a slight deficiency up to $\sim$ 0.3 dex. 
However, this deficiency narrows down at $z=0.5$. 
The abundance of low-mass galaxies ($M_* < 10^{9} \msun$) is sensitive to stellar feedback \citep[e.g., see discussions in][]{Pillepich2018MNRAS.473.4077P}, and the consistency in their abundance implies that the SN wind feedback in the \astrid simulation performs well at regulating galaxy growth. 
In the high-mass regime ($M_*>10^{11} \msun$), galaxy growth is influenced by AGN feedback. 
The consistency in this regime suggests that the AGN feedback model in \astrid results in a reasonable population of massive galaxies. 
We will further investigate the star forming and quiescent galaxy populations in Section~\ref{sec:QG}.

\begin{figure*}[t]
\centering
\includegraphics[width=1.0\textwidth]{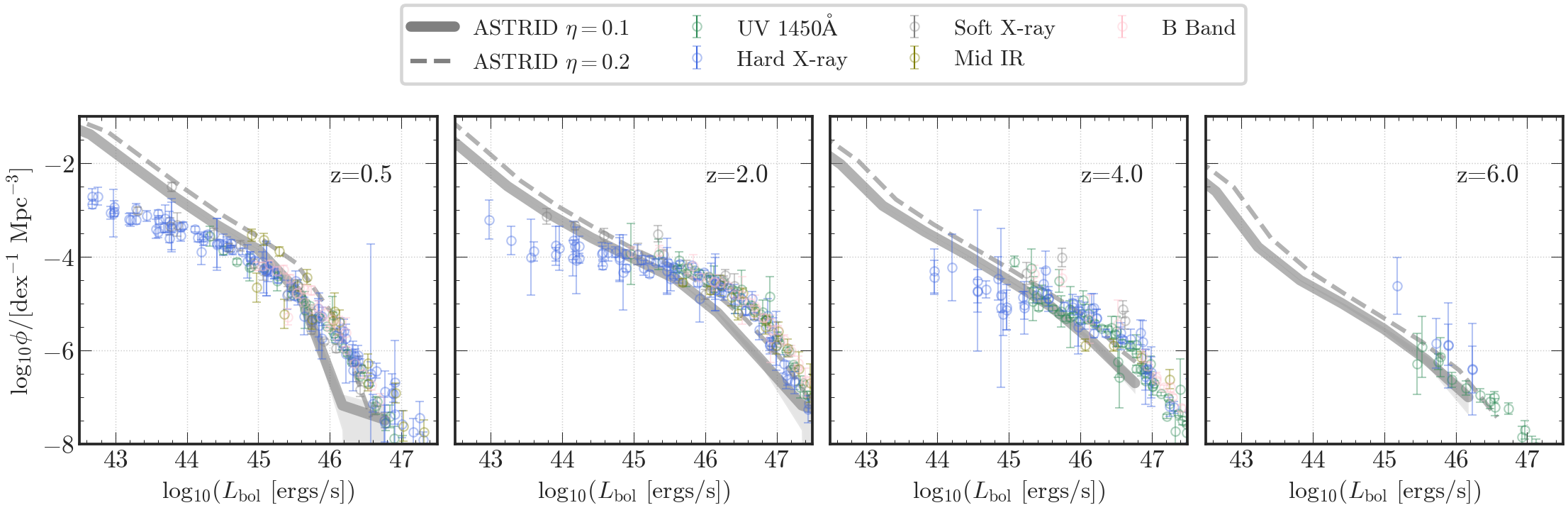}
\caption{Quasar luminosity function of \astrid from $z=0.5$ to $z=6.0$. Colored data points show observational data compiled from \citet{Shen2020MNRAS.495.3252S} (S09). The gray solid lines show the results of BH bolometric luminosity assuming radiative efficiency $\eta=0.1$. The dashed gray lines show the results with radiative efficiency $\eta=0.2$. }
\label{fig:QLF}
\end{figure*}

\begin{figure*}[t]
\centering
\includegraphics[width=1.0\textwidth]{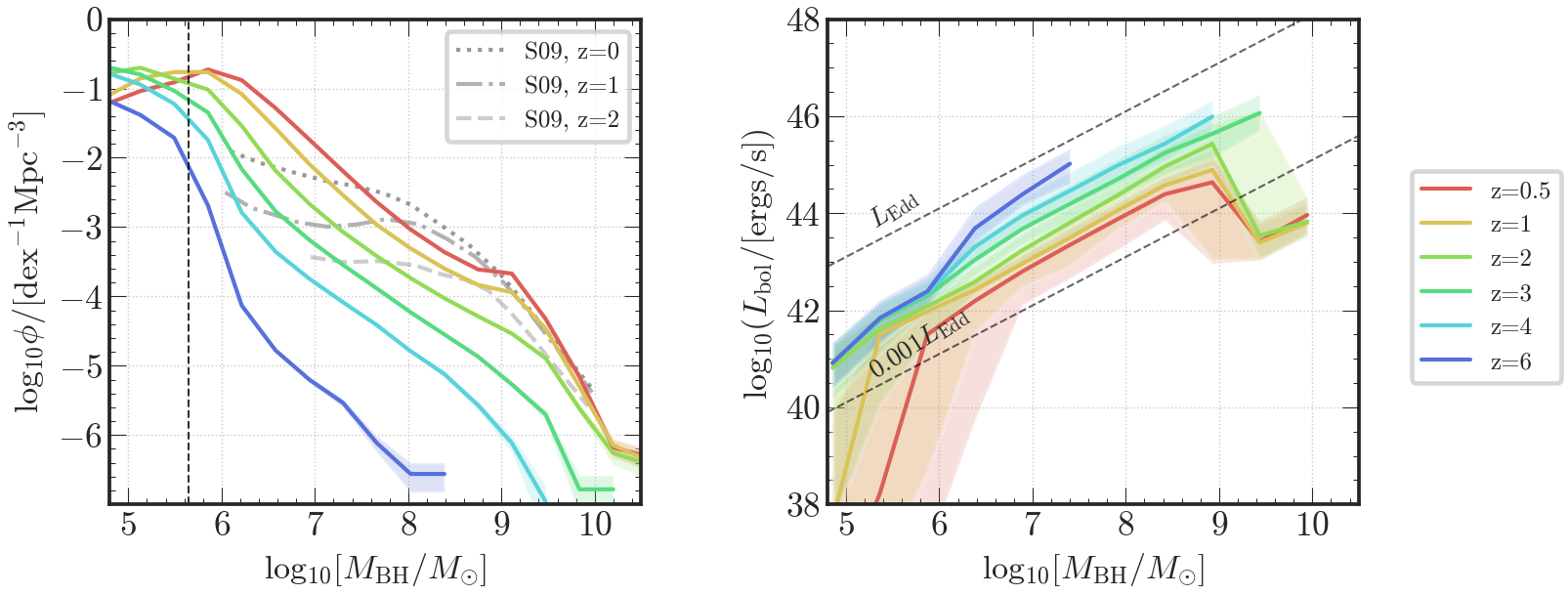}
\caption{
\textit{Left panel}: The colored solid lines show the BH mass function of \astrid from $z=6$ to $z=0.5$. The vertical dashed line marks the BH seed mass of $M_{\rm sd,max} = 3\times 10^{5} \hmsun$. The gray lines show the observational constraint from \citet{Shankar2009ApJ...690...20S}.
\textit{Right panel}: Median bolometric luminosity as a function of BH masses from $z=6$ to $z=0.5$, with the shaded areas indicating the 14-86th percentile of the distributions. The dashed lines mark the Eddington luminosity $L_{\rm Edd}$ and $0.001 \times L_{\rm Edd}$. 
}
\label{fig:bhmf-Lbol}
\end{figure*}

\subsection{Black hole population}
\label{sec3.2:bhmf}

In this section, we examine the global statistics of the BH population in \astridN. 
We present the BH mass function and luminosity function over cosmic time, along with the redshift evolution of the $\mbh - L_{\rm bol}$ relation.

\subsubsection{Black hole luminosity function}

The AGN luminosity function characterizes the population of active BHs. 
It summarizes the comoving space density of AGN as a function of BH luminosity. 
Fig.~\ref{fig:QLF} shows the quasar bolometric luminosity function of \astrid from $z=6$ to $z=0.5$, compared to observational predictions from different bands compiled by \cite{Shen2020MNRAS.495.3252S}.
These observational constraints are obtained by applying different bolometric corrections to the observed luminosity in the associated bands. 
In simulations, the AGN bolometric luminosity is often calculated from the black hole accretion rate with $L_{\rm bol} \propto \epsilon_r \dot{M}_{\rm BH}$, assuming that a given fraction of the accreted mass is converted to light and radiated away.
The choice of radiative efficiency $\epsilon_r$ typically ranges from 10 to 20 per cent in different simulations.
As shown in Eq.~\ref{equation:thermalmode}, the assumption of $\epsilon_r$ is degenerate with the feedback efficiency $\epsilon_{f,th}$, where the product of the two determines the fraction of accreted energy converted to thermal energy.   
The gray solid lines in Fig.~\ref{fig:QLF} show the \astrid AGN bolometric luminosity calculated assuming a uniform radiative efficiency of $\epsilon_r = 0.1$, while the gray dashed lines represent results assuming $\epsilon_r = 0.2$.

\astrid predicts a time evolution of the AGN population qualitatively consistent with the observational trends, with the abundance of bright AGN/quasars peaking at the epoch of cosmic noon around $z=2$ and declining thereafter. 
The AGN luminosity function shows good agreement with observations in the luminosity range of $L_{\rm bol} = 10^{44 \sim 46} \rm ergs/s$ across all redshifts.
At the bright end of $L_{\rm bol} > 10^{46} \rm ergs/s$, \astrid predicts a deficiency of the bright quasars. 
However, we note that the luminosity depends on the assumption of constant radiative efficiency.
A higher radiative efficiency increases the normalization of the luminosity functions and the abundance of bright AGN/quasars. 
As seen in the gray dashed lines of Fig.~\ref{fig:QLF}, the discrepancy in the bright end diminishes with a higher radiative efficiency of $\epsilon_r = 0.2$.
On the other side, \astrid produces an excess of faint AGNs with $L_{\rm bol} < 10^{44} \rm ergs/s$ compared to the hard x-ray measurements. 
Overall, the shape of the \astrid AGN luminosity is steeper than the observational compilations, following the same trend observed in many other simulations \citep[as in, e.g.][for a comparison of an ensemble of cosmological simulations]{Habouzit2022MNRAS.509.3015H}. 

The excess of faint-end AGNs could possibly be due to an overabundance of seed mass BHs, as the halo-based BH seeding prescription might be too optimistic about the prevalence of heavy seed BHs. 
On the other hand, we also note that the detection of faint AGNs is challenging in observations, subject to various limitations such as observational sensitivity, selection effects, and obscuration. 
These factors can result in an underestimation of the true abundance of faint AGNs in observational data.

Quantitative comparison to individual observational datasets would require more sophisticated mock modeling of the simulated BHs and galaxies, accounting for various observational effects such as dust attenuation. 
Despite those challenges, the first-order comparison of the observed quasar luminosity function indicates that \astrid makes reasonable predictions regarding the evolution of AGN abundance over cosmic time.

\subsubsection{Black hole mass function}

As another fundamental property of the black hole population and to assess how it builds up its mass across cosmic history, we show in the left panel of Fig.~\ref{fig:bhmf-Lbol} the \astrid black hole mass function (BHMF) from $z=6$ to $z=0.5$.
The BHMF is calculated as the comoving space density of black holes as a function of the black hole mass. 
The gray lines show the observational inference from \cite{Shankar2009ApJ...690...20S}. 
We see that \astrid predicts overall good agreement with observational results in the mass range $\mbh > 10^7 \msun$. 
We note that there are large uncertainties and inconsistencies with the simulation model in the assumptions applied by these observational inferences of BHMFs. 
Matching to within an order of magnitude across a wide range of $\mbh$ and redshift still validates that the BH model in \astrid reasonably builds up the $\mbh$ population over cosmic time.

In the massive end of $\mbh >10^9 \msun$, the abundance of massive black holes quickly builds up from $z=6$ to $z=2$ and evolves slowly afterward. 
This supports the ``downsizing'' evolution of the black holes, in which more massive black holes build their mass earlier. 
On the other hand, \astrid predicts an excess of the low-mass BH population compared to the shallower shape implied by observations. 
BHs in the mass range $\mbh < 10^{6-7} \msun$ are predominantly seed BHs, as indicated by the vertical line marking $M_{\rm sd,max} = 3 \times 10^5 \hmsun$ ($4.4 \times 10^5 \msun$). 
As discussed in the previous section, while detecting low-mass or faint BHs poses observational challenges, the excess of low-mass BHs predicted by simulations may also be due to the BH seeding prescriptions. 
This highlights the need for future deep X-ray observations to provide more robust constraints on low-mass BHs, which would help to improve the BH seeding models adopted in cosmological simulations.


\subsubsection{$\mbh - L_{\rm bol}$ relation}
\label{subsection3.3}

In the right panel of Fig.~\ref{fig:bhmf-Lbol}, we show the median relation between the bolometric luminosity of the BHs and their masses across different redshifts. 
Most BHs in the mass range of $10^{7-9} \msun$ exhibit bolometric luminosities between 0.1\% and 100\% of the Eddington luminosity. 
For fixed BH masses, the median $L_{\rm bol}$ decreases when going to lower redshift.
Notably, for $z<1$, the accretion rate of low-mass BHs ($\mbh<10^{7}\msun$) drops sharply compared to BHs of the same mass at higher redshifts. 
This redshift evolution is primarily driven by the depletion of the cold dense gas reservoir in galaxies, also known as cosmic starvation.

At $z \leq 2$, the median AGN luminosity declines rapidly for massive BHs ($\mbh > 10^{8.5} \msun$). 
This mass range marks the transition from high-accretion rate AGN thermal feedback, where energy is primarily radiated away in the form of thermal heating, to low-accretion rate AGN kinetic feedback, where energy is deposited mechanically in the form of jets.
The jet-mode kinetic feedback efficiently expels high-density star-forming gas, thereby suppressing the BH accretion for massive BHs.
This process is crucial in regulating the growth of massive BHs and as well as their host galaxies.
As we will show later in Section~\ref{sec3.3:scaling} and Section~\ref{sec:QG}, these BHs in jet mode are also closely associated with suppressed star formation, accounting for the rise of massive quiescent galaxy population after cosmic noon. 

\subsection{Scaling relation between $\mbh$ and $M_*$}
\label{sec3.3:scaling}

\begin{figure*}
\centering
\includegraphics[width=1.0\textwidth]{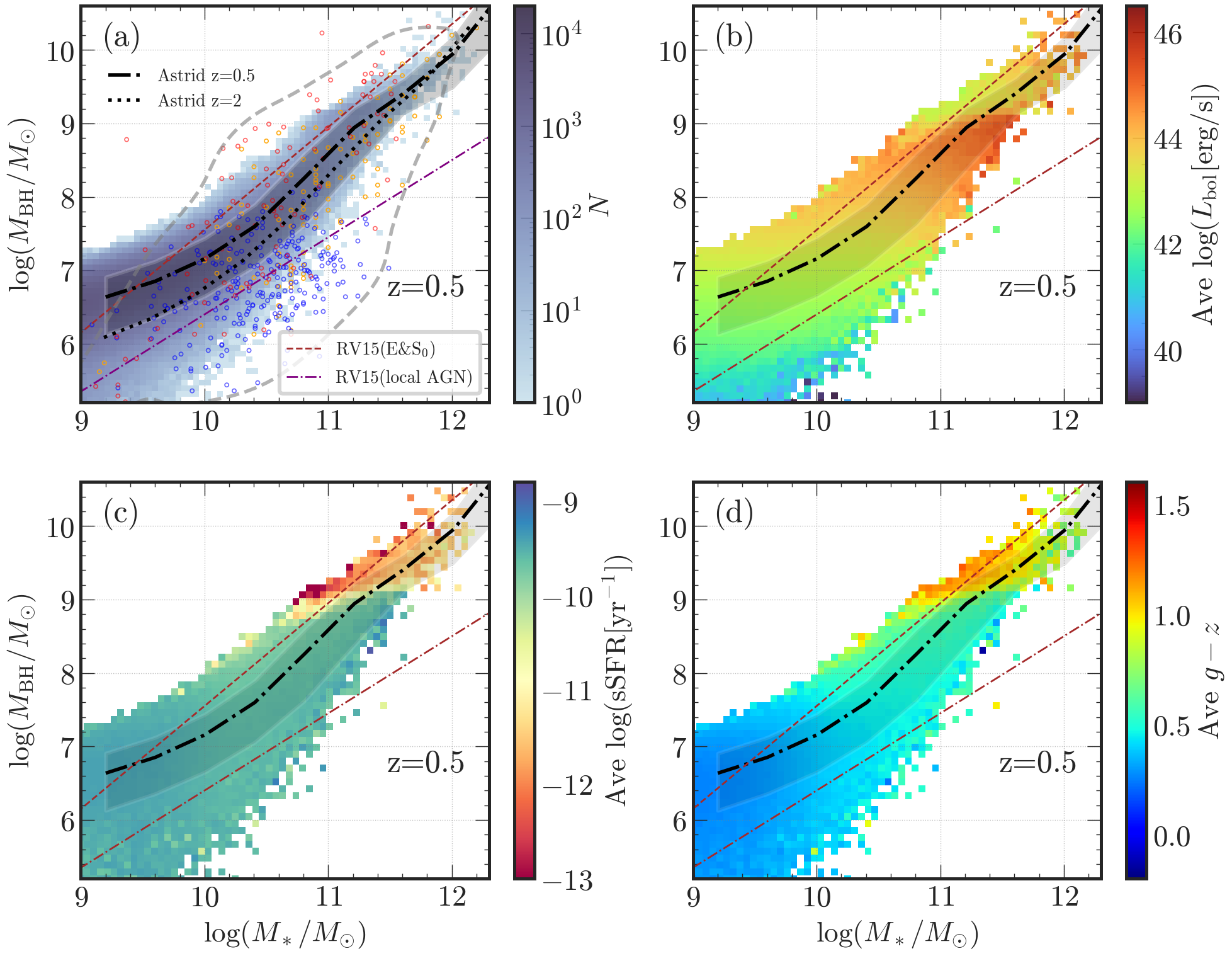}
\caption{
The scaling relation between $M_{\rm BH}$ and $M_*$ in \astridN,  displayed as a 2D histogram in panel (a), is colored by the mean bolometric luminosity of the central SMBH hosted by galaxies in panel (b), the mean specific star formation rate in panel (c), and the mean rest-frame $g-z$ magnitude of the galaxy in panel (d). 
We include only the central galaxies of each FOF group, and the stellar mass $M_*$ is calculated as the stellar mass enclosed within twice the half-mass radius. 
In each panel, the black dashed-dotted line and the gray shaded area represent the median and the 16th-84th percentiles of the overall population, respectively.
We also show the $\mbh-M_*$ relation at $z=2$ in black dotted line in the first panel.   
In panel (a), the colored circles give observational samples from \citet{Reines2015ApJ...813...82R} (blue for their local broad-line AGN samples and red for  early type galaxy samples) and \citet{Graham2023MNRAS.518.2177G} (orange). 
The gray dashed line illustrates the associated KDE contour that encloses 95\% of the provided AGN observational samples. 
To guide the eye, we also show the two $M_{\rm BH}$-$M_*$ relations from \citet{Reines2015ApJ...813...82R}: the upper line corresponds to early type elliptical and spiral/S0 galaxies with classical bulges, and the lower line corresponds to local the broad-line AGN.
}
\label{fig:BH-scaling}
\end{figure*}

\begin{figure*}
\centering
\includegraphics[width=1.0\textwidth]{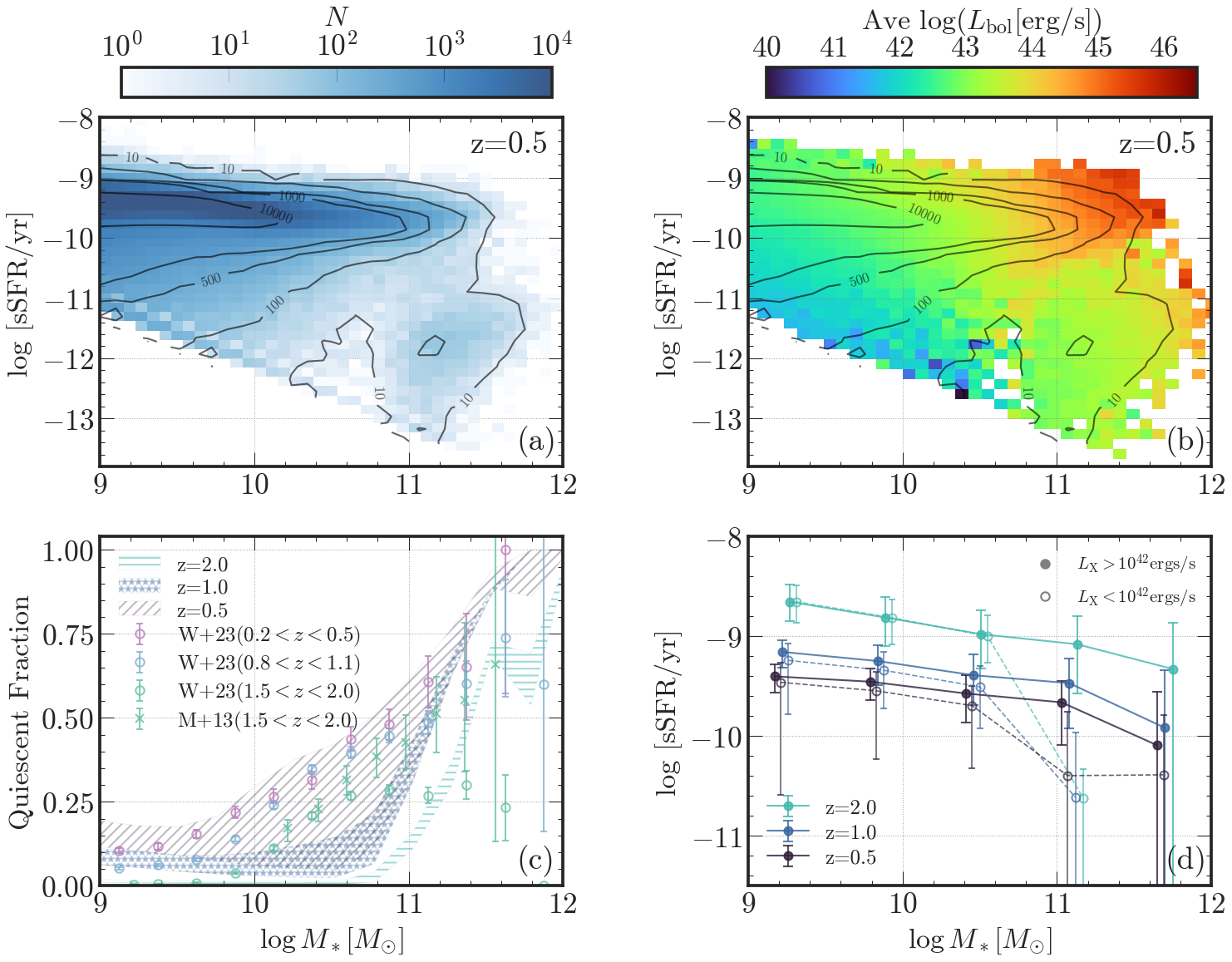}
\caption{
\textit{\bf{Panel (a)}}: 
2D histogram of galaxies in the plane of stellar mass $M_{*}$ and specific star formation rate (sSFR) in \astrid at $z=0.5$. 
The background bluescale histogram shows the number of galaxies as denoted by the black contours. 
For each galaxy, $M_{*}$ is calculated as the stellar mass enclosed within half mass radius of the associated subhalo. 
Star formation rate is calculated as the averaged star formation rate over the past 50 Myrs. 
\textit{\bf{Panel (b)}}:
$M_{*}$ - sSFR relation color-coded by the mean bolometric luminosity of AGN hosted by galaxies in each bin. 
The black contours gives the 2D histogram of the overall population as in panel (a). 
\textit{\bf{Panel (c)}}: 
Quiescent fraction of galaxies as function of the stellar mass. 
The colored hatched regions show the fraction of the quiescent galaxies in \astrid at $z=0.5$, $z=1$ and $z=2$.
We define the quiescent galaxies as galaxies with specific star formation rate $\rm sSFR < sSFR_{thr}$, with $\rm sSFR_{thr}$ ranging from $5\times 10^{-11} \sim 2\times 10^{-10} \mathrm{yr}^{-1}$ to account for uncertainties.
The data points show the observational results compiled in \citet{Weaver2023A&A...677A.184W} (W+23) and \citet{Muzzin2013ApJ...777...18M} (M+13).
\textit{\bf{Panel (d)}}: 
Averaged sSFR as a function of $M_{*}$, for galaxies that host AGN with $L_{\rm X} > 10^{42} \rm ergs/s$ (filled markers) and non-active galaxies (without AGN of $L_{\rm X} > 10^{42} \rm ergs/s$, empty markers).
}
\label{fig:QG-frac}
\end{figure*}

Various empirical scaling relations have been derived between BH mass and different galaxy properties, such as bulge mass and stellar velocity dispersion \citep[e.g., see][]{Kormendy2013ARA&A..51..511K}. 
In this work, we focus on the scaling relation between $\mbh$ and the total stellar mass of the galaxy $M_*$, exploring its correlation with the accretion properties of BHs and star-forming activities. The investigation of scaling relations with other galaxy properties is reserved for future studies.

Fig.~\ref{fig:BH-scaling} presents the $\mbh$-$M_*$ relation colored by various properties, for  galaxies with $M_*>10^{9} \msun$ in \astrid at $z=0.5$.
In this analysis, we include only the central galaxies within each FOF halo. 
For each galaxy, we identify its most massive BH as the central black hole to establish the scaling relation.

Fig.~\ref{fig:BH-scaling} (a) shows the 2D histogram of $\mbh$-$M_*$, with the black line and gray shaded region representing the median and 16-84th percentiles of the overall population. 
This representation is repeated in all four panels.
We note that there is little redshift evolution of the mean $\mbh$-$M_*$ relation from $z=6$ to $z=0.5$, and therefore, we only present the $z=0.5$ results here. 

For observational comparison, we incorporate observational samples from \cite{Reines2015ApJ...813...82R} and \cite{Graham2023MNRAS.518.2177G}, where estimates of the total stellar mass are available. 
We also display a KDE contour that encloses a large fraction of the AGN observational samples to illustrate the scatter. 
To guide the eye, we add scaling relations between $\mbh$ and $M_*$ derived in \cite{Reines2015ApJ...813...82R} for both inactive samples of early-type elliptical galaxies and active broad-line AGN.

We can see that \astrid exhibits a relatively tight scaling relation (at the massive end) between $\mbh$-$M_*$, with the median relation lying between the fits for inactive and active galaxies, as expected. 
It also shows a fair level of scatter, populating the observed broad-line AGN region of $\mbh \sim 10^{6-8} \msun$ and $M_{*}=10^{10.5-11} \msun$. 
This region is often not well-reproduced by many simulations \citep[see,e.g.][]{Habouzit2021MNRAS.503.1940H}. 

We further investigate the $\mbh$-$M_*$ relation with AGN and star-forming activities.
Panels (b), (c) of Fig.~\ref{fig:BH-scaling} show the $\mbh$-$M_*$ relation colored by the mean AGN bolometric luminosity (in log scale) and the specific star formation rate ($s$SFR) in each bins. 
We use $s$SFR, defined as the star formation rate (SFR) normalized by the galaxy's stellar mass, to quantify the relative intensity of star formation activity.
For each galaxy in \astridN, we calculate their rest-frame magnitudes in the SDSS $g$, $r$, $i$, and $z$ bands based on SED modeling using the \textsc{fsps} package \citep{Conroy2010ascl.soft10043C}. 
Panel (d) of Fig.~\ref{fig:BH-scaling} is colored by the rest-frame $g-z$ color magnitude of the galaxies, with larger $g-z$ values representing redder galaxies primarily composed of older stellar populations.


Many observational studies have shown that early-type inactive galaxies host more massive BHs compared to late-type galaxies with active AGN \citep[e.g.,][]{Reines2015ApJ...813...82R,Greene2020ARA&A..58..257G}. This trend is qualitatively well reproduced by \astridN. As shown in panel (c), for massive central galaxies with $M_* > 10^{10.5} \msun$, those with more massive BHs of $\mbh>10^{8.5} \msun$ that reside above the global scaling relation tend to be quenched, with an average suppressed specific star formation rate of $s\rm SFR < 10^{-11} \rm{yr}^{-1}$. 
As can be seen in panel (b), these massive BHs also have lower luminosities compared to their lower-mass counterparts in galaxies of the same stellar mass due to strong self-regulation via AGN feedback.

The jet-mode AGN kinetic feedback activates when BHs grow beyond a mass threshold of $5\times10^8 \msun$, efficiently expelling gas from the galaxies, curtailing BH gas accretion, and resulting in the cessation of star formation. 
On the most massive BH end, i.e., for BHs with $\mbh > 10^{9} \msun$ residing in galaxies with $M_* >10^{11} \msun$, the simulated host galaxies are very red, with $g-z$ colors greater than 1. 
Conversely, for lower-mass BHs with $\mbh<10^{8} \msun$ in $M_* \sim 10^{10} \msun$ galaxies, the majority of host galaxies are blue and star-forming, with $g-z < 0.5$.

In the intermediate mass range of the $\mbh-M_*$ relation, where $\mbh = 10^{8-9} \msun$ and $M_* \sim 10^{10-11} \msun$, the host galaxies exhibit a mix of red quenched and blue star-forming galaxies, resulting in an average $g-z \sim 0.7$. 
Interestingly, the BHs in this regime also have the most efficient accretion and therefore the brightest AGN luminosities, likely undergoing a transition phase from star formation to quiescence.

\begin{figure*}
\centering
\includegraphics[width=1.0\textwidth]{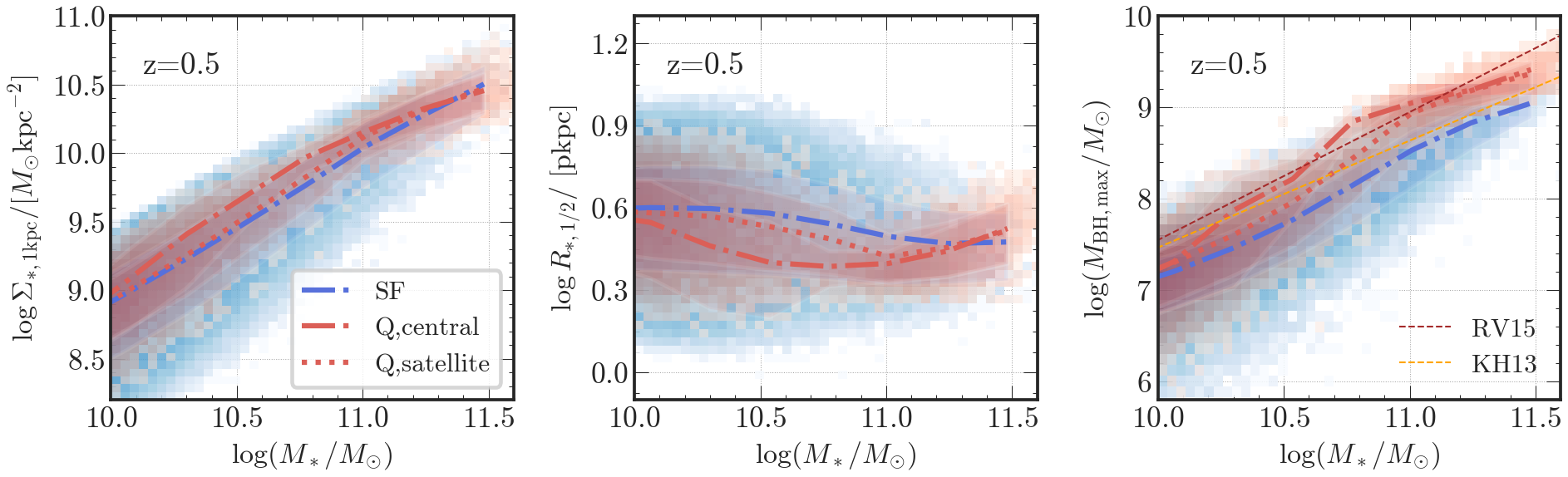}
\caption{
The relationship between galaxy mass and their compactness, size, and central black hole mass. 
The $x$-axis in each panel represents the galaxy mass, defined as the stellar mass enclosed within twice the half-mass radius.
The $y$-axis of the left panel shows the stellar surface density within 1 kpc, the middle panel shows the half-mass radius of the galaxy, and the right panel gives the mass of the most massive black hole within the galaxy. The blue and red regions represent the 2D histograms of star-forming and quiescent galaxy populations, respectively, distinguished by the quenching criterion $s\rm{sfr} < 5 \times 10^{-11} \rm yr^{-1}$. 
The blue and red shaded regions indicate the 16th to 84th percentiles for both star-forming and quiescent populations. 
The blue dashed-dotted lines give the median of the star-forming galaxies. 
Within the quiescent galaxy population, galaxies are further categorized as either central or satellite galaxies. 
The red dashed-dotted lines represent the median of quenched central galaxies, while the red dotted lines show the median of quenched satellite galaxies.
}
\label{fig:Mstar-Sigma1-Rhalf}
\end{figure*}


\section{Quiescent galaxies}
\label{sec:QG}

In this section, we study the population of quiescent galaxies in \astridN. 
In Section~\ref{subsection:fraction-QG}, we calculate the quiescent fraction of galaxies across cosmic time and compare it to observations. 
Section~\ref{subsection:AGN-QG} examines the correlation between AGN activity and quiescent galaxies on a statistical basis. 
Section~\ref{subsection:size-QG} shows the size and compactness of the quiescent galaxy population compared to star-forming galaxies. 
In Section~\ref{subsection:timescale-QG}, we investigate the correlation between quenching timescales and galaxy morphological properties, such as galaxy size and age gradient. 
Finally, in Section~\ref{subsection:z4-QG}, we briefly show the quenching scenarios for galaxies at $z>4$.


\subsection{Fraction of Quiescent Galaxies}
\label{subsection:fraction-QG}

We start by examining the time evolution of quiescent galaxies in \astrid in Fig.~\ref{fig:QG-frac}. 
Panel (a) shows the relationship between $s$SFR and galaxy stellar mass. 
The SFR of each galaxy is averaged over the past 50 Myrs based on the distribution of stellar ages. 
This definition is chosen because it closely correlates with observational tracers like the H$\alpha$ emission line, which is used as an indicator of recent star formation \citep{Kennicutt1998ARA&A..36..189K}.
We have tested alternative definitions, such as the instantaneous SFR (based on the gas particles) within twice the half-mass radius, and found that using different definitions does not lead to qualitative changes in the results.

A common observational method to identify quiescent galaxies is through color-color diagrams, such as the rest-frame UVJ diagram \citep[e.g.,][]{Muzzin2013ApJ...777...18M}. 
However, simulations require more refined modeling of dust attenuation in certain color bands to accurately replicate these diagnostics. 
Instead of rest-frame UVJ colors, we classify non-star-forming (or passive) galaxies based on their $s$SFR.
One typical threshold for $s$SFR in non-local, higher redshift galaxies is $s\rm SFR = 1/[2t_{univ}(z)]$, where $t_{\rm univ}(z)$ denotes the age of the universe \citep[as used in, e.g.,][]{Tacchella2022ApJ, Park2023ApJ...953..119P}.
This corresponds to $s\rm SFR = 5\times 10^{-11} \rm yr^{-1}$ for $z=0.5$ and $s\rm SFR = 2 \times 10^{-10} \rm yr^{-1}$ for $z=2$.

The calculation of quiescent fraction can be sensitive to the applied $s\rm SFR$ thresholds. 
To account for uncertainties, we use the aforementioned $s\rm SFR$ thresholds as upper and lower bounds to calculate the QG fraction as a function of galaxy stellar mass across redshifts $z=2$ to $z=0.5$ in panel (c). 
The hatched region represents the range of quiescent fractions calculated between these two $s$SFR thresholds.
For observational comparisons, the colored data points correspond to quiescent fraction results from the COSMOS2020 galaxy survey catalog, compiled by \citet{Weaver2023A&A...677A.184W}. 
Observational separation of star-forming and quiescent galaxies based on photometric colors becomes increasingly challenging at higher redshifts. 
As a complimentary comparison, we also include quiescent fraction result at $z\sim2$ compiled by \citet{Muzzin2013ApJ...777...18M}, which is based on different sample selections and photometric color.

Overall, \astrid predicts the quiescent galaxy population qualitatively consistent with observational results. 
The fraction of quiescent galaxies increases at lower redshifts (as indicated in panel (d)) and with increasing galaxy mass. 
Beyond $M_* > 10^{10.5} \msun$, a significant portion of galaxies exhibit suppressed star formation. 
As illustrated in Fig.~\ref{fig:BH-scaling}, the population of massive quenched galaxies corresponds to systems with overmassive black holes that lie above the overall $\mbh-M_*$ scaling relation. 
These central black holes exhibit suppressed AGN luminosity in a radiatively inefficient regime and operate in a kinetic AGN feedback mode, which effectively quenches star formation in these massive galaxies.

We note that the quiescent galaxy fraction in the mass range $M_* = 10^{10-11} \msun$ is lower than observational constraints, particularly at high redshift ($z=2$). 
This discrepancy is likely due to AGN feedback being insufficiently efficient in this mass range. This aligns with the challenge faced by many simulations (including \astrid), which tend to underpredict the quiescent galaxy abundance in the early universe \citep[see,e.g.,][for $z\geq3$ results]{Weller2024arXiv240602664W}.
In the next subsection, we explore the statistical correlation between observed AGN activity and star formation. 

\subsection{AGN and star formation activity}
\label{subsection:AGN-QG}

To assess how AGN activity and its feedback affect star formation, one approach is to systematically compare star formation activity between AGN hosts and controlled samples of inactive galaxies (without detectable AGN).
Numerous observational studies have explored differences in star formation activity between AGN hosts and inactive galaxies. 
Recent observations suggest that AGN hosts generally exhibit comparable levels of star formation activity to inactive galaxies of similar mass and redshift. 
Some studies indicate slightly lower SFR in AGN hosts \citep[e.g.,][]{Bongiorno2012MNRAS.427.3103B, Mullaney2015MNRAS.453L..83M}, while others report slight increases \citep[see, e.g.,][]{Juneau2013ApJ...764..176J, Cowley2016}. 
Despite variations in approach and selection techniques across these studies, the overall consensus suggests that star formation activity in AGN hosts aligns closely with that of normal, inactive galaxies. 
Consequently, some studies question the significant role of AGN feedback as a primary mechanism for galaxy quenching \citep[e.g.,][]{Cowley2016, Mahoro2017}.


We have illustrated in Fig.~\ref{fig:BH-scaling} that, in \astridN, the AGN feedback is the primarily quenching mechanisms for massive galaxies ($M_* > 10^{10.5} \msun$). 
In Fig.~\ref{fig:QG-frac}, panel (b) gives the mean AGN bolometric luminosity in the $s$SFR-$M_*$ plane, while panel (d) presents the averaged $s$SFR as a function of $M_*$ for both AGN host galaxies and inactive galaxies.
Here we classify AGN host galaxies using the criterion of AGN luminosity $L_X > 10^{42}$ ergs/s \citep[as applied in, e.g.,][]{Cowley2016}. 
That corresponds to $L_{\rm bol} \sim 2\times10^{43}$ ergs/s assuming an X-ray bolometric correction from \cite{Hopkins2007}.

Panel (b) of Fig.~\ref{fig:QG-frac} illustrates the two competing factors contributing to the observed correlation between SFR and AGN activity.
Firstly, AGN activity and star formation in galaxies often share a common triggering mechanism, both fueled by the same reservoir of cold gas. Consequently, bright AGN hosts tend to be more gas-rich and also exhibit higher levels of star formation, leading to a positive correlation between AGN accretion and star formation rates.
Secondly, AGN activity can suppress star formation. 
The population of massive, quenched galaxies in the lower right corner of panel (b) is attributed to efficient AGN feedback, which depletes the cold, star-forming gas within the galaxy. 
However, the strong self-regulation of AGN feedback can also result in low AGN luminosity, reducing the likelihood of detecting the SMBH as active.
Considering these competing effects, the overall trend observed in panel (d) shows that, on average, AGN hosts exhibit a slight elevation in star formation activity compared to inactive galaxies across all redshifts from $z=2$ to $z=0.5$. 
This elevation is more pronounced at higher stellar masses of $M_* > 10^{10.5} \msun$, as quenched galaxies in this mass regime typically host radiatively inefficient BHs operating in jet mode feedback, resulting in lower X-ray luminosity and making these SMBH less detectable.

Our prediction aligns with observational findings indicating that the star formation activity of X-ray AGN hosts is generally consistent, with a slight elevation compared to normal inactive galaxies. 
However, it is important to note that \textit{this trend does not indicate positive AGN feedback on star formation}. 
Instead, it arises from two opposing factors: a positive correlation between AGN activity and star formation, as both are fueled by a shared gas reservoir, and a negative effect where AGN feedback ultimately quenches star formation.


\begin{figure*}
\centering
\includegraphics[width=1.0\textwidth]{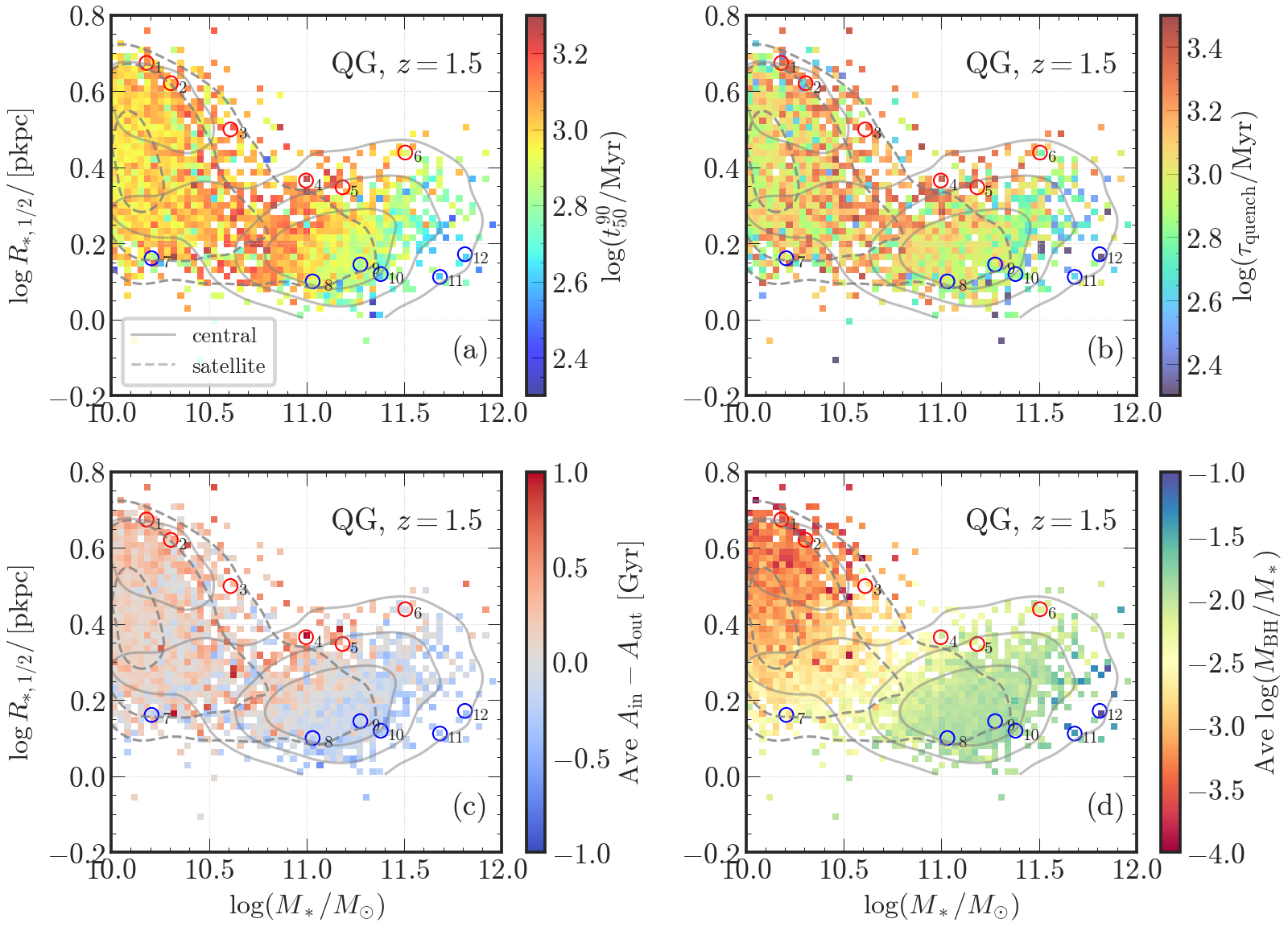}
\caption{
Relationship between the galaxy stellar mass and size for the quiescent galaxy population in \astrid at $z=1.5$. 
Quenched galaxies are defined by the quenching criterion  $s\rm{SFR} < 1/2 t_{\rm univ}(z)$ (corresponding to $1.2 \times 10^{-10} \rm yr^{-1}$ at $z=1.5$). 
The gray solid and dashed lines show the KDE contours for the central and satellite quenched galaxies separately. 
The four panels show the 2D histograms colored by the star formation timescale $t^{90}_{50}$ (panel a), quench timescale (panel b), radial age difference $A_{\rm in} - A_{\rm out}$ (panel c), and the ratio between the central black hole mass and the stellar mass (panel d).
}
\label{fig:QG-Mstar-size-t59}
\end{figure*}

\begin{figure*}
\centering
\includegraphics[width=1.0\textwidth]{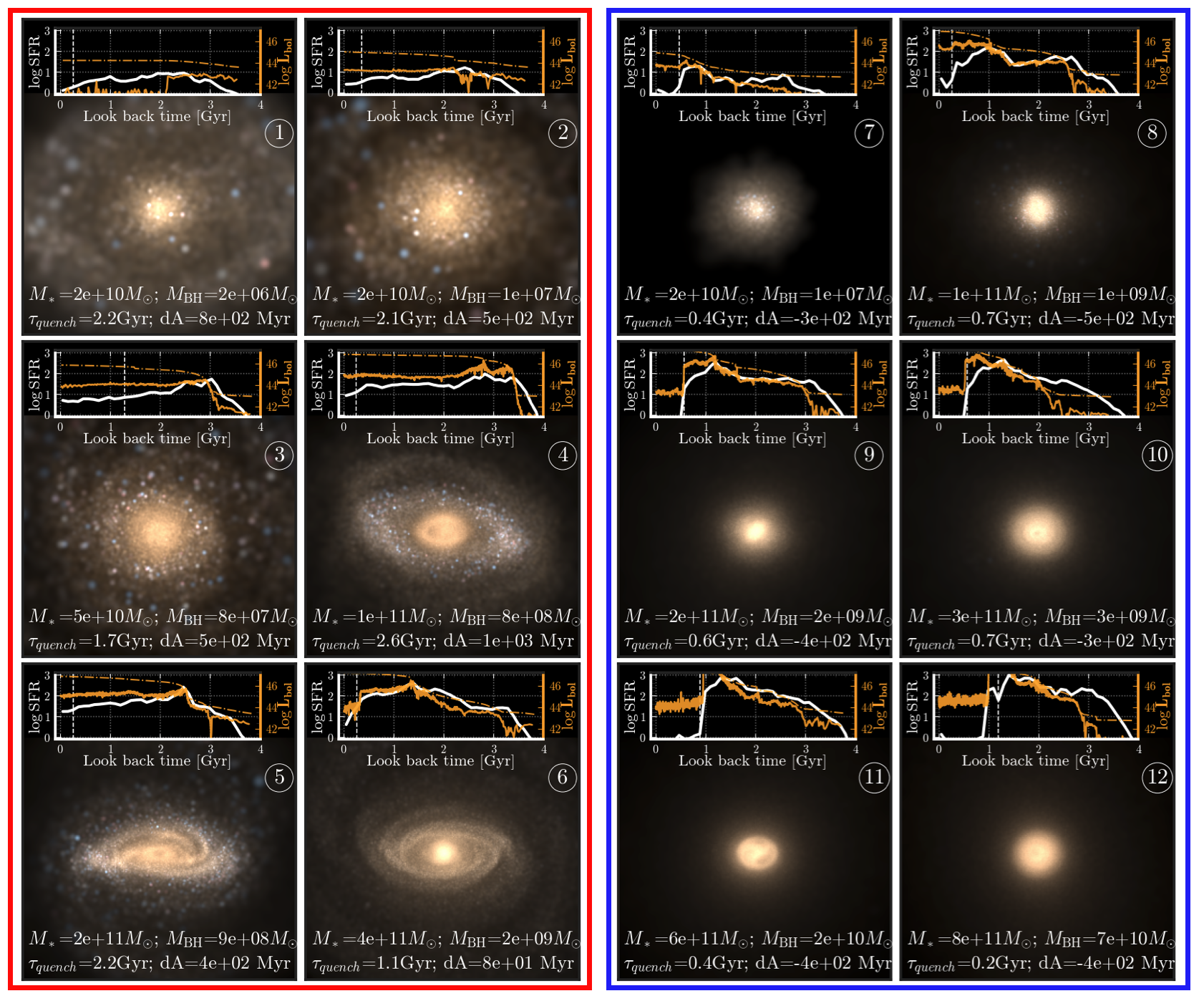}
\caption{
We select 6 examples of diffuse quiescent galaxies (marked as red circles in Fig.~\ref{fig:QG-Mstar-size-t59}) and compact quiescent galaxies (marked as blue circles) and plot their mock images. 
Each panel shows a 30 $\hkpc$ region with mock images, where the RGB channels represent the flux in the rest-frame \textit{grz} bands. 
The upper insets display the star formation history of the galaxies (white line) as a function of look-back time, along with the light curve evolution of the central BH (orange line). The orange dashed-dotted line indicates the Eddington luminosity of the BH at that time. The vertical white dotted line indicates when the galaxy became quenched.
}
\label{fig:QG-Mstar-size-t59-example}
\end{figure*}

\subsection{Size and compactness of the quiescent galaxies}
\label{subsection:size-QG}
 
Different quenching mechanisms can leave distinct morphological imprints to galaxies. 
Many studies use the central surface density of stars within 1 kpc ($\Sigma_{*,1\rm kpc}$) to quantify galaxy compactness and find it to be a powerful predictor of quiescence \citep[e.g.,][]{Barro2016ApJ,Suess2021ApJ...915...87S}.
These studies suggest that galaxies undergoing central compaction-like mechanisms, where dissipative processes funnel gas into the central region and trigger starburst, tend to quench rapidly and exhibit more compact morphologies compared to star-forming galaxies and occupy the higher region of the $\Sigma_{*,1\rm kpc} - M_*$ plane \citep[e.g.,][]{Woo2019MNRAS.487.1927W}.

In this section, we explore the correlations between galaxy quiescence and galaxy morphology/central density. 
Fig.~\ref{fig:Mstar-Sigma1-Rhalf} shows the morphological properties of quiescent galaxies, focusing on their size and compactness to investigate differences between star-forming and quiescent galaxy populations.
The first panel shows the $\Sigma_{*,1\rm kpc}$-mass relation. Galaxies located in higher regions of the $\Sigma_{*,1\rm kpc} - M_*$ plane are considered more compact.
The second panel shows the galaxy size-mass relation, with $R_{*,1/2}$ representing the half-mass radius of the stellar component for each galaxy. 
We present histograms of the $\Sigma_{,1\rm kpc} - M_*$ relation and the $R_{*,1/2} - M_*$ relation for both star-forming (blue) and quiescent galaxies (red). 
For the QG population, we further divide them into central and satellite galaxies and plot their median relations.

We observe a clear increase in compactness for central quenched galaxies in the mass range $M_* = 10^{10.5-11} \msun$, with higher $\Sigma_{*,1\rm kpc}$ and smaller $R_{*,1/2}$ compared to star-forming galaxies of the same mass. 
Meanwhile, satellite galaxies only show a slight increase in compactness compared to the star-forming population.
As shown in the third panel of Fig.~\ref{fig:Mstar-Sigma1-Rhalf}, within the QG population, central galaxies on average host more massive BHs, which are overmassive compared to the overall $\mbh - M_*$ relation. 
These overmassive black holes are fueled by the intense gas inflow that results in the compact galaxy morphology, while simultaneously exerting efficient AGN feedback that quenches the galaxy. 
In contrast, satellite galaxies mainly reside in massive halos and are subject to external quenching mechanisms on larger spatial scales unrelated to physical processes in the galaxy center, thus not exhibiting a significant difference in size and compactness compared to the overall galaxy population of the same mass.

\subsection{Relation between the quenching timescale and galaxy morphology}
\label{subsection:timescale-QG}

Advances in high-resolution spectroscopic and integral field unit (IFU) observations are enhancing our understanding of the evolutionary pathways of quenching by revealing details about the star formation history and morphological features of quiescent galaxies, including stellar age profiles, color gradients, galaxy size and compactness.
Several observational studies indicate that different quenching mechanisms can lead to distinct morphological features and quenching timescales  \citep[see, e.g.,][]{Barro2016ApJ, Carnall2019, Woo2019MNRAS.487.1927W, Suess2021ApJ...915...87S}.
Generally, galaxies undergoing central compaction-like quenching mechanisms--more common at high redshifts--tend to quench rapidly, are more compact, and exhibit younger galaxy centers (i.e., positive color gradients). 
In contrast, galaxies that quench over longer timescales typically have larger sizes and older central regions (negative color gradients).

Recent spatially resolved spectroscopic observations are starting to unveil the morphological details, such as size and color profile, of quiescent galaxies in higher redshift regimes of $z>1$ \cite[e.g.,][]{Nelson2021MNRAS.508..219N,Suess2021ApJ...915...87S}.
In these high-redshift regimes, a larger fraction of galaxies undergo rapid quenching, often on timescales as short as a few hundred Myrs, such as the post-starburst galaxies \citep[e.g.,][]{Park2023ApJ...953..119P}. 
In this section, we present the QG population at $z=1.5$ in \astridN, examining the correlation between quenching timescales and morphological properties to provide theoretical understanding and make predictions for future observations.

Fig.~\ref{fig:QG-Mstar-size-t59} presents the galaxy size-mass relation for all quiescent galaxies at $z=1.5$.  
Each pixel in the 2D histograms is color-coded to illustrate the correlation between quenching properties and galaxy morphological features. 
In panel (a), the color represents the mean star formation timescale, $t^{90}_{50}$, defined as the time required for galaxies to build between the 50th and 90th percentile of their stellar mass. 
Panel (b) is color-coded by the quenching timescale, $\tau_{\rm quench}$, calculated as $t_{\rm SF peak} - t_{\rm quench}$, where $t_{\rm peak}$ is the epoch when star formation reach its peak, and $t_{\rm quench}$ is the time of the quenching epoch when the galaxy has the $s$SFR drops below $s$SFR$=1/[2t_{\rm univ}(z)]$.
The \astrid QG population displays a wide range of quenching timescales, consistent with observations of QGs at $z > 1.5$, where quenching timescales vary significantly, from approximately 100 Myr to over 1 Gyr \citep[e.g.,][]{Tacchella2015Sci...348..314T,Belli2019ApJ...874...17B}. 
In panel (c), the color indicates the galaxy age (or color) gradient, represented by the radial age difference, $A_{\rm in} - A_{\rm out}$, which is the difference between the mean age of stars within the half-mass radius and those outside of it. Finally, panel (d) is color-coded by the average $\mbh / M_*$ ratio.
Each panel in Fig.~\ref{fig:QG-Mstar-size-t59} uses Gaussian KDE contours to show the distributions of central and satellite quenched galaxies.
We can see that massive, compact quenched galaxies ($M_* > 10^{10.5} \msun$) are predominantly central galaxies, whereas lower-mass quiescent galaxies are mainly satellites and tend to have more diffuse morphologies with larger size.

To illustrate some example QGs with compact and diffuse morphologies (marked as blue and red circles in in Fig.~\ref{fig:QG-Mstar-size-t59}), Fig.~\ref{fig:QG-Mstar-size-t59-example} shows mock galaxy images in the rest-frame $grz$ color bands, with the upper insets displaying the star formation history and light curve evolution of the central BH as a function of look-back time.


As shown in panel (c) of Fig.~\ref{fig:QG-Mstar-size-t59}, among massive QGs with $M_* > 10^{10.5} \msun$, those with larger sizes (larger $R_{*,1/2}$) tend to have older stellar populations in their centers compared to their outskirts.
These galaxies also experience longer timescales for both star formation and quenching, typically on the order of $>1$ Gyr. 
Fig.~\ref{fig:QG-Mstar-size-t59-example} panel $1-6$ provides examples of these diffuse QGs, spanning a wide mass range from $10^{10} \msun$ satellite galaxies to massive central galaxies.
These QG systems consistently exhibit older stellar populations in the central regions and younger stellar populations in the outskirts. 
Panel (d) shows that many diffuse QGs have lower $\mbh / M_*$ ratios compared to other QGs of the same mass, suggesting that AGN feedback may play a less dominant role in the quenching process, especially in lower-mass galaxies.
The time evolution of the BH light curve and star formation history indicates that most of these systems evolve passively, with the central BH maintaining a steady quasar (thermal) mode and minimal variability, while star formation gradually declines over several Gyrs.
The quenching in these galaxies is primarily driven by the suppression of gas inflow at the galactic or circumgalactic scale, gradually cutting off the `wet inflow' (i.e., gas inflow rate $>$ SFR) that sustains star formation in the galaxy center. 
This quenching process exhibits an `inside-out' pattern, with star formation in the outskirts being quenched later than in the center, resulting in older central regions and younger outskirts.

On the other hand, quenched galaxies with more compact sizes (smaller $R_{*,1/2}$) typically have shorter timescales for both star formation and quenching. 
As shown in Fig.~\ref{fig:QG-Mstar-size-t59-example} panel $7-12$, many of these QG systems undergo a phase of intense starburst shortly before rapid quenching.
These galaxies generally have younger stellar populations in their centers compared to their outskirts and exhibit higher $\mbh / M_*$ ratios than quenched galaxies of similar mass, suggesting that AGN feedback is a critical factor in their quenching.
The evolution of the BH light curve reveals that during the starburst phase, the BH often accretes near the Eddington limit. 
In many cases (e.g., examples 9-12), the rapid growth of BH mass, combined with thermal heating from quasar-mode feedback, drives the BH into a radiatively inefficient regime, activating AGN jet-mode feedback.
The jet-mode feedback rapidly quenches the galaxy by ejecting star-forming gas, reducing the star formation rate to below $1 \msun$/yr, after which the galaxy evolves passively in a maintenance mode.
This process explains the `post-starburst galaxies' in the high-$z$ universe, where dissipative processes drive intense gas inflow toward the central region, triggering both bursts of star formation and black hole growth, as well as morphological transformation into compact spheroids.
These galaxies build the bulk of their stellar mass over a short timescale in the central region and undergo rapid quenching due to efficient AGN kinetic feedback.
This scenario is consistent with recent observational evidence suggesting that kinetic (ejective) AGN feedback is responsible for quenching post-starburst galaxies \citep[e.g.,][]{Eugenio2023arXiv230806317D,Belli2024Natur.630...54B}.


In short, \astrid predicts a clear correlation between quenching timescales and morphological properties of galaxy size and age gradient. 
As proposed by several observational studies \citep{Woo2019MNRAS.487.1927W, Suess2021ApJ...915...87S}, rapidly quenching galaxies are often associated with central or internal mechanisms, particularly the AGN feedback. 
These galaxies typically undergo a compaction-like core-building event that fuels the central AGN, triggering efficient AGN feedback and leading to rapid quenching. 
They tend to have more compact morphologies compared to galaxies of similar mass and feature younger stellar populations in their centers relative to their outskirts. 
On the other hand, galaxies that are quenched over longer quenching timescales usually have more extended morphologies and are characterized by older central stellar populations, exhibiting an `inside-out' quenching pattern.

\subsection{$z>4$ quenched galaxies}
\label{subsection:z4-QG}

\begin{figure}
\centering
\includegraphics[width=1.0\columnwidth]{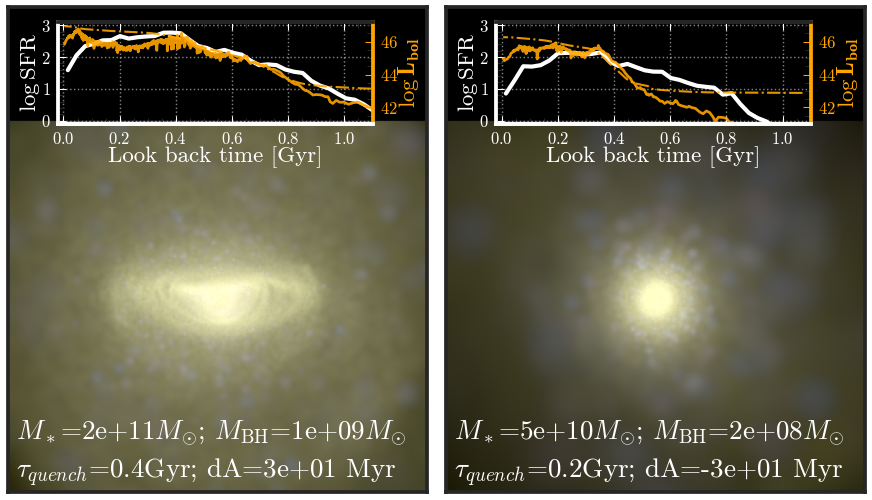}
\caption{
Illustration of two example quenched galaxies at $z=4$.
Each panel displays a 30 $\hkpc$ region, with mock images generated using the JWST observed F090W, F277W, and F444W bands for the RGB channels. 
The upper insets in each panel show the star formation history of the galaxies (white line) as a function of look-back time, alongside the light curve evolution of the central black hole (orange line). The orange dashed-dotted line represents the Eddington luminosity of the BH at that time.
}
\label{fig:QG-high-z}
\end{figure}

Recent JWST observations have uncovered a surprisingly large population of quiescent galaxies at higher redshifts of $z>3$, indicating that galaxy quenching processes were active within the first few billion years of the universe \citep[e.g.,][]{Carnall2023, Long2023, Valentino2023}. 
On the simulation side, however, several studies \citep[e.g.,][]{Merlin2019MNRAS.490.3309M,Weller2024arXiv240602664W} have pointed out that many cosmological simulations, including \astridN, significantly underestimate the abundance of $z>3$ quenched galaxies compared to recent high-redshift observational constraints.  
This discrepancy suggests the need to reassess the timescales and physical mechanisms driving galaxy quiescence.

In \astridN, AGN feedback operates solely in thermal mode at redshifts $z>2.3$. 
Yet, as reported by \citet{Weller2024arXiv240602664W}, \astrid does produce a small population of quenched galaxies in this high-redshift regime, albeit with much lower number densities.
As presented in \citet{Weller2024arXiv240602664W}, these high-$z$ quenched massive galaxies tend to host overmassive BHs compared to the overall $\mbh-M_{*}$ scaling relation, suggesting that AGN feedback plays a crucial role in quenching those galaxies. 
In Fig.~\ref{fig:QG-high-z}, we present two examples of such high-$z$ quenched systems at $z=4$. 
The quenching timescales for galaxies at higher redshifts are systematically shorter due to the younger age of the universe ($t_{\rm age}(z=4) \sim 1.5$ Gyr). 
As shown in the evolution of the BH light curve and star formation history, the star formation rate begins to gradually decline after the central BH reaches peak accretion (near the Eddington limit), operating in quasar mode with a luminosity of $L_{\rm bol} \gtrsim 10^{46}$ erg/s.
This implies that the thermal feedback from the central massive BH can still quench star formation by heating the surrounding gas.

Although the number of quiescent galaxies in this high-$z$ regime is too limited in \astrid to conduct statistical analysis, the observed examples display a qualitatively similar trend as the quiescent galaxies at lower redshifts.
Specifically, galaxies with more extended morphologies (left panel) tend to have relatively longer quenching timescales and older stellar centers compared to more compact ones (right panel). 
This suggests that the correlation between galaxy morphology and quiescence, as discussed in Section~\ref{subsection:timescale-QG}, may persist despite the difference in detailed AGN feedback prescriptions at high redshifts.

\section{Summary and conclusion}
\label{Sec:summary}

This work presents new results from the \astrid simulation spanning the redshift range from $z=3$ to $z=0.5$, covering the period around and after the epoch of cosmic noon.
In Section~\ref{Sec:galaxy-bh-population}, we compare the galaxy and BH populations with observational data and examine the $\mbh$-$M_*$ scaling relation, along with its correlations with BH accretion and star-forming activities. The key findings are summarized as follows:

\begin{itemize}

\item The galaxy stellar mass function (GSMF) in \astrid generally aligns well with observational data across redshifts $z > 0.5$ at both the low-mass and high-mass ends. However, in the `knee' region, around $M_* \sim 10^{10.5-11.2} \msun$, there is a slight deficiency of up to 0.3 dex at $z=2$ and $z=1$, but this discrepancy diminishes by $z=0.5$.

\item 
The black hole mass function in \astrid matches observational data for $\mbh > 10^7 \msun$, indicating a rapid buildup of massive black holes from $z=6$ to around $z \sim 2$, followed by slower evolution, consistent with the observed `downsizing' trends.
The AGN luminosity function from \astrid simulations matches observations well in the luminosity range of $L_{\rm bol} = 10^{44 \sim 46} \rm ergs/s$ across all redshifts.
However, it predicts fewer bright quasars with $L_{\rm bol} > 10^{46} \rm ergs/s$ compared to observed data. This discrepancy might be reconciled by assuming a higher radiative efficiency, $\eta = 0.2$.
On the other hand, like many other cosmological simulations, \astrid exhibits an excess of faint AGNs (about 1 dex higher at $L_{\rm bol} \sim 10^{43} \ \rm ergs/s$) compared to hard X-rays observations.

\item  
\astrid exhibits $\mbh$-$M_*$ scaling relations consistent with empirical observations across all redshifts. It also reproduces a diversity of systems with scatters that are comparable to the observational range.
This scaling relation, correlated with star formation and AGN activity, reveals that massive BHs ($\mbh \gtrsim 10^{9} \msun$) in galaxies with $M_* > 10^{10.5} \msun$ are in very red, quenched systems, while lower-mass BHs in less massive galaxies are in blue, actively star-forming environments, indicating the significant role of AGN feedback in quenching galaxies.

\end{itemize}

As the population of quiescent galaxies rises after the epoch of cosmic noon, we conduct a detailed investigation of the \astrid quiescent galaxies in Section~\ref{sec:QG}.
The main findings are summarized below.

\begin{itemize}
\item We present the fraction of quiescent galaxies in \astrid across cosmic time, showing that the quiescent fraction increases at lower redshifts and among more massive galaxies. These trends show qualitative agreement with observational results from $z = 2$ to $z = 0.5$.

\item We analyze the relationship between AGN activity and star formation rates. 
Although AGN feedback is crucial in quenching star formation in massive galaxies ($M_* > 10^{10.5} \msun$), AGN activity and star formation remain statistically positively correlated due to their shared cold gas reservoir.
In \astridN, the AGN-host galaxies on average exhibit a slight elevation in star formation activity compared to inactive galaxies across $z=2$ to $z=0.5$.
This trend aligns with observational findings indicating that the star formation activity of X-ray AGN hosts is generally consistent, with a slight elevation compared to normal inactive galaxies.

\item We investigate the morphological properties of quiescent galaxies in \astridN.
Central galaxies, primarily quenched by AGN feedback, tend to host more massive black holes and exhibit more compact morphologies compared to their star-forming counterparts. In contrast, satellite quenched galaxies show less distinct differences in size, compactness, and central black hole mass, as they are predominantly quenched by environmental effects without significant involvement of AGN feedback.

\item 
Different quenching mechanisms leave distinct morphological imprints on quenched galaxies. In \astridN, massive, compact quiescent galaxies typically undergo shorter quenching timescales, have younger central regions, and host overmassive black holes. This is often the result of a compaction-like quenching mechanism that funnels gas into the galaxy center, leading to both starbursts and rapid BH growth, which in turn triggers AGN jet-mode feedback that rapidly quenches the galaxy. 
On the other hand, quiescent galaxies with more diffuse morphologies generally have older stellar populations in their central regions compared to the outskirts, with AGN feedback typically playing a less significant role in the quenching process. 
These galaxies typically experience longer quenching timescales due to quenching processes that gradually deplete the galactic-scale star-forming gas.

\end{itemize}


\section*{Acknowledgements}
We would thank Katherine Suess and Arianna Long for helpful discussions. 
YN acknowledges support from the ITC Postdoctoral Fellowship.
DM acknowledges support from NASA grant 80NSSC22K0722.
TDM and RACC acknowledge funding from  the NSF AI Institute: Physics of the Future, NSF PHY-2020295, NASA ATP NNX17AK56G, and NASA ATP 80NSSC18K101. 
TDM acknowledges additional support from  NSF ACI-1614853, NSF AST-1616168, NASA ATP 19-ATP19-0084, NASA ATP 80NSSC20K0519, and RACC from NSF AST-1909193.
SB acknowledges funding from NASA ATP 80NSSC22K1897.

\bibliographystyle{aasjournal}
\bibliography{example} 





\end{document}